\documentclass[11pt,a4paper]{article}
\usepackage[dvips]{graphicx}
\usepackage{jcappub}
\usepackage{enumitem}

\usepackage{color}

\title{Interacting warm dark matter}

\author[a]{Norman Cruz,}
\author[a]{Guillermo Palma,}
\author[a]{David Zambrano}
\author[b]{and Arturo Avelino}

\affiliation[a]{Departamento de F\'isica, Facultad de Ciencia,\\
Universidad de Santiago de Chile,\\
Casilla 307, Santiago, Chile.}

\affiliation[b]{Departamento de F\'isica, DCI, Campus Le\'on, \\
Universidad de Guanajuato, \\
CP. 37150, Le\'on, Guanajuato, M\'exico.}

\emailAdd{norman.cruz@usach.cl}

\abstract{We explore a cosmological model composed by a dark matter fluid interacting with a dark energy fluid. The interaction term has the non-linear $\lambda \rho_{\rm m}^{\alpha} \rho_{\rm e}^{\beta}$ form, where $\rho_m$ and $\rho_e$ are the energy densities of the dark matter and dark energy, respectively. The parameters $\alpha$ and $\beta$ are in principle not constrained to take any particular values, and were estimated from observations. We perform an analytical study of the evolution equations, finding the fixed points and their stability properties in order to characterize suitable physical regions in the phase space of the dark matter and dark energy densities. The constants ($\lambda, \alpha, \beta$) as well as $w_{\rm m}$ and $w_{\rm e}$ of the EoS of dark matter and dark energy respectively, were estimated using the cosmological observations of the type Ia supernovae and the Hubble expansion rate $H(z)$ data sets.  We find that the best estimated values for the free parameters of the model correspond to a warm dark matter interacting with a phantom dark energy component, with a well goodness-of-fit to data. However, using the Bayesian Information Criterion (BIC) we find that this model is overcame by a warm dark matter -- phantom dark energy model without interaction, as well as by the $\Lambda$CDM model. We find also a large dispersion on the best estimated values of the $(\lambda, \alpha, \beta)$ parameters, so even if we are not able to set strong constraints on their values, given the goodness-of-fit to data of the model, we find that a large variety of theirs values are well compatible with the observational data used.}


\arxivnumber{}

\begin{document}
\maketitle

\flushbottom

%%%%%%%%%%%%%%%%%%%%%%%%%%%%%%%%%%%%%%%%%%%%%%%%%%%%%%%%%%%%%%%%%%%%%%%%%%%%%%%%%%%%%%%
\section{Introduction}\label{intro}
%%%%%%%%%%%%%%%%%%%%%%%%%%%%%%%%%%%%%%%%%%%%%%%%%%%%%%%%%%%%%%%%%%%%%%%%%%%%%%%%%%%%%%%

The existence of a dark component with an exotic equation of state,
i. e., with a  ratio $w =  p/\rho$ negative and close to $-1$, which
drives an accelerated expansion, is consistent with data coming from
type Ia Supernovae (SNe Ia)~\cite{SNe}, large scale structure
formation (LSS)~\cite{LSS}, cosmic microwave background radiation
(CMB)~\cite{CMB}, baryon acoustic oscillations (BAO)~\cite{BAO} and
weak lensing~\cite{lensing}.

Cosmic observations show that densities of dark energy (DE) and dark
matter (DM) are of the same order today, despite their different
decreasing rates. To solve the so--called coincidence
problem~\cite{coincidence} an evolving dark energy field with a
non-gravitational interaction with matter~\cite{Amendola} is proposed
(decay of dark energy to dark matter). In this case both dark fluids
interact via an additional coupling term in the fluid equations. In
the current research different forms of coupling have been
considered. Most of them study the coupling between cold DM and DE.
In general, the interactions investigated are particular cases the
form $\lambda_{m} H \rho_m + \lambda_{e} H \rho_{e}$, where $H$ is
the Hubble parameter, and $\rho_m$, $\rho_{e}$ are the dark matter
and dark energy densities respectively~\cite{Interactions}.

Nevertheless, non linear interactions of the form $\lambda \frac{
\rho_m \rho_{e}}{\rho_m + \rho_{e}}$ were considered in~\cite{Cai}.
A more plausible interaction is inspired by the situation of two
types of fluids interacting, where the interaction is proportional
to the product of the powers of the energy density of both
components. In this case, the interaction rate goes to zero as one or
both densities become zero, and increases when each of the densities
grows. We then consider in this paper a general interaction of the form
$\lambda\rho_m^\alpha\rho_e^\beta$, where the parameter $\lambda$
has dimensions of $[energy\text{-}density^{\alpha +\beta -1}\times time]^{-1}$.

This type of interaction was investigated in the framework of an
holographic dark energy~\cite{Ma} and in coupled
quintessence~\cite{Mangano,Laura-Etal-2010}, where the evolution of the energy
densities of the dark interacting components was investigated for
different values of the parameters $\alpha$ and $\beta$. A cyclic
scenario for the present situation $\rho_{e}\sim\rho_{m}$ was found
as a possible solution to the coincidence problem. The particular
case $\alpha=\beta=1$ was studied in~\cite{Lip}. In this case, if energy is transferred from dark energy to dark matter ($\lambda
> 0$), and for phantom type dark energy ($w_{e}<-1$), the
energy densities also display periodic orbits. A general result,
which is independent of the interaction type, and only assumes
that the energy is transferred from dark energy to dark matter,
pointed out that stationary solutions for the ratio
$r=\rho_{m}/\rho_{e}$ require a phantom dark
energy~\cite{Arevalo}.

Nevertheless, these stationary solutions do not guarantee a solution
of the coincidence problem. For example, in~\cite{X.Chen} a DE
modelled by a phantom field was studied in the framework of
interaction terms proportional to a density. The results of this
investigation indicates that in all cases the late-time solutions
correspond to a complete DE domination and as a consequence the
coincidence problem remains unsolved. A suitable coupling, with the
form $\frac{\rho_{e}\rho_{m}}{\rho_{e}+\rho_{m}}$, was chosen
in~\cite{Z.Guo} for a phantom field evolving under an exponential
potential. An accelerated late-time phase with a stationary ratio of
the energies densities of the two components was found.

The requirement of phantom DE to obtain stationary solutions can be
seen as a kind of theoretical need for this type of matter in the
framework of an interacting dark sector. From the observational
point of view, phantom DE is supported by new SN Ia
data~\cite{Alam}~\cite{Corasaniti}, analysis of the cosmic microwave
background (CMB) and large-scale structure. This model is also
preferred by WMAP data combined with either SNe Ia or baryon
acoustic oscillations (BAO). Nevertheless, from the theoretical
point, fluids that violate the null energy condition $\rho +p \geq
0$ present problems, such as UV quantum instabilities of the
vacuum~\cite{Carrol}. For a phantom DE with constant EoS parameter
$w_e$, its the late time behavior is characterized by an increasing
energy density which becomes infinite in a finite time (Big Rip).
The interaction of a phantom fluid with dark matter can prevent this
type of behavior, leading to finite energy densities during the
cosmic evolution and avoiding of future singularities.

The aim of this paper is to study the time evolution of the dark
sector densities when the interaction mentioned above is considered.
This evolution is driven by a highly non-linear coupled differential
equations when the parameters $\alpha$ and $\beta$ are left free. We
solve numerically these equations and also study analytically the
stability of the fixed-points. Since it has been argued that the second
law of thermodynamics and Le Ch\^atelier's principle implies
$\lambda > 0$~\cite{Pavon} we do not consider here the case of
energy transferred from dark matter to dark energy.

In most of the previous investigation which take into account this
type of non lineal interaction, cold dark matter with zero pressure
is taken to be interacting with the dark energy fluid. Nevertheless,
current research has opened the possibility that a warm dark matter
component fits better new results found at the level of galaxies and
cluster of galaxies~\cite{WDMworks}. For a wide discussion on this
matter see~\cite{workshop}.

When both dark fluids are under interaction, their effective EoS
behaves following Eq. (\ref{wmeff}) (see below in Section II). For
cold dark matter,  the effective EoS always corresponds to an exotic
fluid with negative pressure, nevertheless for warm dark matter the
effective EoS parameter could change the sign during the cosmic
evolution. Since up to date we do not know which is the nature of
this matter, the proposed model together with observations allow
either exotic dark matter or warm dark matter. An interacting warm
dark matter allows these two possibilities and so it assumes from
the beginning a positive EoS constant parameter for this component.

In this work we investigate how a warm dark matter modifies the
behavior of an interacting dark sector and the constraints for its
EoS (assuming a barotropic form) derived from astrophysical
observations.

Our paper is organized as follows. In section \ref{dark} we present the
effective equations of state for the interacting dark components,
assuming that the energy is transferred from dark energy to dark
matter. We also extend the result found in~\cite{Arevalo} in order
to find the condition for stationary points of the parameter $r$
when the dark matter is assumed not to be dust.  In section \ref{stability} we
study the two coupled differential equations corresponding to
continuity equations of both interacting fluids. We study
analytically the fixed-points including their stability. In section
\ref{numerical}, we obtain numerical solutions using a numerical method of
adaptive step-size algorithm called Bulirsch-Stoer method. We
explore the behavior of the fixed points varying the parameters
$\alpha$, $\beta$ and the EoS of both dark sector components. In
section \ref{constraints} we use result from cosmological observation to find the
best values for the free parameters, $( \alpha, \beta,
\bar{\lambda}, w_{\rm m})$, of the corresponding theoretical model.
Finally, in section \ref{conclusions} the main results are summarized and
different physical scenarios consistent with cosmological
observations are discussed.

%%%%%%%%%%%%%%%%%%%%%%%%%%%%%%%%%%%%%%%%%%%%%%%%%%%%%%%%%%%%%%%%%%%%%%%%%%%%%%%%%%%%%%%
\section{General considerations of an interacting dark sector}\label{dark}
%%%%%%%%%%%%%%%%%%%%%%%%%%%%%%%%%%%%%%%%%%%%%%%%%%%%%%%%%%%%%%%%%%%%%%%%%%%%%%%%%%%%%%%

In the following we assume a flat FRW universe filled basically with the
fluids of the dark sector. We consider a warm dark matter of density $\rho_m$
and a dark energy component described by the density $\rho_e$. For simplicity
we also assume that both fluids obey a barotropic EoS, so we have $p_m = w_m
\rho_m$ for the warm dark matter and $p_e = w_e \rho_e$ for the dark energy. In
what follows we restrict our model to the late time of cosmic evolution, which
implies that the others components of the universe, like radiation and baryons
are negligible. In this case the sourced Friedmann equation is given by
\begin{eqnarray}
  3 H^2 & = & \rho_{m} + \rho_e,
  \label{Frid}
\end{eqnarray}
where $8\pi G=1$ has been adopted. We will assume that the dark matter component is interacting with the dark energy component, so their continuity equations take the form
\begin{eqnarray}
\dot{\rho}_m + 3(1 + w_m) H \rho_m & = & + Q \label{rhom}\\
\dot{\rho}_e + 3(1 + w_e) H \rho_e& = &  - Q\text{,}\label{rhoe}
%  \dot{\rho}_{b} & = & - 3 H \rho_b \\
%  \dot{\rho}_{r} & = & - 4 H \rho_r \\
 \end{eqnarray}
where $H = \dot{a} / a$ is the Hubble parameter, and $a (t)$ is the scale-factor. Here, an overdot indicates a time derivative. $Q$ represents the interaction term, despite that we do not use a specific functional dependence at this stage, we will only assume that $Q$ do not change its sign during the cosmic evolution.

\subsection{Effective EoS for interacting fluids}\label{effective_EoS}

Let us discuss briefly the behavior of the dark components in terms of an effective EoS driven by the interacting term. Rewriting the continuity equation (\ref{rhom}) in the usual form
\begin{eqnarray}
\dot{\rho}_m + 3(1 + w_{m_{eff}}) H \rho_m& = & 0 ,\label{rhomeff}
\end{eqnarray}
where $w_{m_{eff}}$ represents the effective EoS for the interacting dark matter, which is given by
\begin{eqnarray}
w_{m_{eff}} & = & w_{m}-\frac{Q}{3H\rho_m}.\label{wmeff}
\end{eqnarray}

Note that the behavior of $w_{m_{eff}}$ can be quite different if $w_{m}$ is non zero. With the usual assumption of cold dark matter ($w_m=0$) and $Q>0$, which implies that energy is transferred from dark energy to dark matter, some kind of exotic dark matter with a negative EoS is driven, assuming, of course, that we are in an expanding universe $H>0$. For warm dark matter we would have, depending on the type of interaction considered and the strength of the coupling constant appearing in $Q$, a possible change of the sign on the effective EoS during the cosmic evolution. For the dark energy, the effective EoS is given by
\begin{eqnarray}
w_{\Lambda_{eff}} & = & w_e+\frac{Q}{3H\rho_e}.\label{weeff}
\end{eqnarray}

For $Q>0$ the above equation indicates, even for $w_e=-1$ (cosmological constant), that the effective Dark Energy (DE) fluid will behave as a quintessence field. Then, an effective phantom behavior can only be obtained if $w_e<-1$.

\subsection{The behavior of the coincidence parameter $r$}\label{coincidence}

In order to address the coincidence problem in terms of the dynamics of the parameter $r\equiv \rho_m/\rho_e$, we will make in what follows a similar analysis along the line describes in ref.~\cite{Arevalo}. The dynamics of the parameter $r$ is given by
\begin{eqnarray}\label{requation}
\dot{r}&=& r \left( \frac{\dot{\rho_{m}}}{\rho_{m}} - \frac{\dot{\rho_e}}{\rho_e} \right)\text{,}
\end{eqnarray}
where the dot indicates derivative with respect to the cosmic time. Changing the time derivatives by a derivative with respect to $\ln a^{3}$, which will be denoted by a prime, i.e. $\dot{\rho}=\rho^{\prime}3H$, eqs (\ref{rhom}) and (\ref{rhoe}) go into
\begin{eqnarray}
\frac{\rho^{\prime}_{m}}{\rho_{m}} =-(1 + w_{m}) + \frac{Q}{3 H
\rho_m },  \,\,\, \,\,\, \frac{\rho^{\prime}_e}{\rho_e} =-(1 +
w_e) - \frac{Q}{3 H \rho_e }\text{.}
\end{eqnarray}

In terms of total density $\rho =\rho_{e} + \rho_{m}$, we obtain
\begin{eqnarray}\label{rhoprime}
\rho^{\prime} =  -   \left [   1 +  \frac{w_{m} r +
w_{e}}{r+1}\right ] \rho.
\end{eqnarray}

For the coincidence parameter $r$ the evolution equation reads
\begin{eqnarray}\label{rprime}
r^{\prime} =  r \left [ (w_e-w_m)  +
\frac{Q}{3H}\frac{(1+r)^{2}}{r\rho}\right ]\text{.}
\end{eqnarray}

Before specifying any particular type of interaction, we will
discuss some general properties of these equations. The critical
point of Eq. (\ref{rhoprime}) is given by
\begin{eqnarray}\label{rcritic}
r_{c}  =  - \frac{(1 + w_{e})}{w_{m} +1}.
\end{eqnarray}

If $w_{m}>0$ (warm dark matter) and since $r$ must be positive, it follows that $w_{e} <-1$, which corresponds to a phantom DE. Using the condition $r^{\prime} = 0$ in Eq.(\ref{rprime}) and the value of $r_{c}$ one finds \begin{eqnarray}\label{rhocritic}
\rho_{c}  =  - \frac{(w_{m}-w_{e})}{w_{m} +1}~\frac{Q}{3H(1 +
w_{e})}\text{.}
\end{eqnarray}

Depending on the value of $w_{m}\geq 0$ there are two interesting cases.

\subsection{Case $ w_{m}=0$}\label{wm0}
In this case, Eq. (\ref{rhocritic}) simplifies to
\begin{eqnarray}\label{casea}
\rho_{c}  =   \frac{w_{e}}{w_{e} +1}~\frac{Q}{3H}\text{.}
\end{eqnarray}

This case was analyzed in~\cite{Arevalo}, concluding that a positive
stationary energy density $\rho_{c}$ requires $Q >0$ since $w_{e} <-1$.  This
means that independent on the interaction type, the existence of critical
points requires a positive exchange from dark energy to dark matter (DM).

\subsection{Case $w_{m} > 0$ (warm dark matter)}\label{warm_dark}

For this case the expression for $\rho_{c}$ becomes
\begin{eqnarray}\label{caseb}
\rho_{c}  =   \frac{w_{e}-w_{m}}{w_{m} +1}\frac{Q}{3H(1 + w_{e})}\text{.}
\end{eqnarray}

The condition $w_{e} <-1$ leads in this case to the same result
found in section \ref{wm0} for the sign of $Q$. This result also
holds for $-1<w_{m}<0$.

We find here the general condition to have accelerated expansion
in terms of the energy densities of the dark components and their
EoS. Differentiating (\ref{Frid}) with respect to $t$ and
substituting for $\dot{\rho}_m$ and $\dot{\rho}_{e_{}}$ gives the
auxiliary equation
\begin{equation}
  2 \dot{H} = - (1 + w_{m}) \rho_m - (1 + w_e) \rho_e\text{.}\label{Hub2}
\end{equation}

The acceleration is given by the relation $\ddot{a} = a ( \dot{H}
+ H^2)$. From (\ref{Frid}) and (\ref{Hub2}), we obtain
\begin{equation}
  \ddot{a} = - \frac{a}{6} (\rho_m (1 + 3 w_{m})+ (1 + 3 w_{e}) \rho_e) . \label{ACCEL}
\end{equation}

The condition  $\ddot{a} >0$ leads to the inequality
\begin{equation}
\rho_e > - \frac{1 + 3 w_{m}}{1 + 3 w_{e}} \rho_m .
\label{ConditionACCEL}
\end{equation}

Since we shall consider $w_{e}<-1$ this condition represent a
right line with positive slope $1 + 3 w_{m}/ 3|w_{e}|-1$ in the
plane $(\rho_m,\rho_e)$.

%GP and DZ
%%%%%%%%%%%%%%%%%%%%%%%%%%%%%%%%%%%%%%%%%%%%%%%%%%%%%%%%%%%%%%%%%%%%%%%%%%%%%%%%%%%%%%%
\section{Evolution equations: Fixed points and stability analysis}\label{stability}
%%%%%%%%%%%%%%%%%%%%%%%%%%%%%%%%%%%%%%%%%%%%%%%%%%%%%%%%%%%%%%%%%%%%%%%%%%%%%%%%%%%%%%%

Introducing the interaction term $Q=\lambda
\rho_m^{\alpha}\rho_{e}^{\beta}$ in Eqs. (\ref{rhom}) and
(\ref{rhoe}) yields
\begin{eqnarray}
  \dot{\rho}_m & = & - 3(1 + w_{m}) H \rho_m + \lambda \rho_m^{\alpha} \rho_{e}^{\beta} \label{rhom2} \\
  \dot{\rho}_e & = & - 3(1 + w_e) H \rho_e - \lambda \rho_m^{\alpha} \rho_e^{\beta}\text{.} \label{rhoe2}
%  \dot{\rho}_{b} & = & - 3 H \rho_b \\
%  \dot{\rho}_{r} & = & - 4 H \rho_r \\
 \end{eqnarray}

The time evolution of the dark matter and dark energy densities is
given by the highly non-linear coupled differential equations
(\ref{rhom2}) and (\ref{rhoe2}). We rewrite them as follows:
\begin{equation}
    \dot{\rho}_{_i} = f_{_i}(\rho_m,\rho_e)\text{,}
    \label{dynamics_rho}
\end{equation}
where $i=m,e$. The functions $f_m$ and $f_e$ are defined by the following expressions
\begin{eqnarray}
f_m(\rho_m,\rho_e) & = & - \sqrt{3}\left(1 + w_m\right)\rho_m\left(\rho_m+\rho_e\right)^{1/2} + \lambda \rho_m^\alpha \rho_e^\beta \label{ICF1} \\
f_e(\rho_m,\rho_e) & = & - \sqrt{3}\left(1 +
w_e\right)\rho_e\left(\rho_m+\rho_e\right)^{1/2} - \lambda
\rho_m^\alpha \rho_e^\beta\text{.}
\label{ICF2}
\end{eqnarray}

From numerical results we expect that the above equations have some
non-trivial fixed-points $\left(\bar{\rho}_m,\,\bar{\rho}_e\right)$,
which we want to study analytically including their stability
properties. In spite of the non-linearities and according to ref.
\cite{Elsgoltz}, it is still possible to analyze the stability of fixed
points by using the linearized piece of the original differential
equations
\begin{equation}
\dot{\rho}_{_i} = \sum_j a_{ij} \rho_j + R_{i}(\rho_m,\rho_e)
\label{dynamics_rho2}
\end{equation}
where $R_{i}(\rho_m,\rho_e)$ includes all the non linearities of the original system of equations (\ref{ICF1}) and (\ref{ICF2}), provided the inequality
\begin{equation}
|R_{i}(\rho_m,\rho_e)| \leq N\left( \sum_i \rho_i^2
\right)^{\frac{1}{2} + \gamma} \label{elsgoltz_ineq}
\end{equation}
is fulfilled in a neighbour region around the fixed points, for some positive constants $N$ and $\gamma$. To achieve this goal we expand both the dark matter and dark energy
densities around their fixed point values $\left(\bar{\rho}_m,\,
\bar{\rho}_e\right)$ as follows:

\begin{eqnarray*}
  \rho_m^\alpha = \left(\bar{\rho}_m + \mu\rho_m\right)^\alpha & = & \bar{\rho}_m^{\,\,\,\alpha} + \alpha\mu\rho_m + \mathcal{O}\left(\mu^2\rho_m^2/\bar{\rho}_m\right) \\
    \rho_e^\beta = \left(\bar{\rho}_e + \mu\rho_e\right)^\beta & = & \bar{\rho}_e^{\,\,\,\beta} + \beta\mu\rho_e + \mathcal{O}\left(\mu^2\rho_e^2/\bar{\rho}_e\right)\text{,}
\end{eqnarray*}
it also holds

\begin{equation*}
    \left(\rho_m+\rho_e\right)^{1/2} = \left(\bar{\rho}_m^{\,\,\,\alpha} + \bar{\rho}_e^{\,\,\,\beta}\right)^{1/2} + \frac{1}{2\left(\bar{\rho}_m^{\,\,\,\alpha} + \bar{\rho}_e^{\,\,\,\beta}\right)^{1/2}} \mu \left(\rho_m+\rho_e\right) + \mathcal{O}\left(\mu^2\right).
\end{equation*}

Inserting the above perturbative expressions into the differential
equation system one obtains up to first order in $\mu$:
\begin{equation}
\dot{\rho}_{_i} = \sum_j a_{ij} \rho_j
%\label{dynamics_rho2} % DUPLICADA !
\end{equation}
or explicitly
\begin{eqnarray}
    \dot{\rho}_m &=& -\sqrt{3}\left(1+\omega_m\right) \left[\frac{\bar{\rho}_m\left(\rho_m+\rho_e\right)}{2\left(\bar{\rho}_m+\bar{\rho}_e\right)^{1/2}} +\rho_m\left(\bar{\rho}_m+\bar{\rho}_e\right)^{1/2}\right] + \lambda\left[\beta\rho_e\bar{\rho}_m^{\,\,\,\alpha} + \alpha\rho_m\bar{\rho}_e^{\,\,\,\beta} \right] \label{LSE1}\\
    \dot{\rho}_e &=& -\sqrt{3}\left(1+\omega_e\right) \left[\frac{\bar{\rho}_e\left(\rho_m+\rho_e\right)}{2\left(\bar{\rho}_m+\bar{\rho}_e\right)^{1/2}} +\rho_e\left(\bar{\rho}_m+\bar{\rho}_e\right)^{1/2}\right] - \lambda\left[\beta\rho_e\bar{\rho}_m^{\,\,\,\alpha} + \alpha\rho_m\bar{\rho}_e^{\,\,\,\beta} \right]\text{.} \label{LSE2}
\end{eqnarray}

The tree level values are implicitly defined by the relations
\begin{eqnarray}
  \sqrt{3}\left(1 + w_m\right)\bar{\rho}_m\left(\bar{\rho}_m+\bar{\rho}_e\right)^{1/2} & = & \lambda \bar{\rho}_m^{\,\,\,\alpha} \bar{\rho}_e^{\,\,\,\beta} \label{TLE1}\\
  \sqrt{3}\left(1 + w_e\right)\bar{\rho}_e\left(\bar{\rho}_m+\bar{\rho}_e\right)^{1/2} & = & -\lambda \bar{\rho}_m^{\,\,\,\alpha} \bar{\rho}_e^{\,\,\,\beta}
  \label{TLE2},
\end{eqnarray}
or equivalently,
\begin{eqnarray}
    \bar{\rho}_e &=& -\frac{1+\omega_m}{1+\omega_e} \label{Fix_P_m}\\
    \bar{\rho}_m &=& \left[ (-1)^\beta\frac{\lambda}{\sqrt{3}}\frac{(1+\omega_m)^{\beta-1}(1+\omega_e)^{1/2-\beta}}{(\omega_e-\omega_m)^{1/2}} \right]^{(3/2-\alpha-\beta)^{-1}}\text{.} \label{Fix_P_e}
\end{eqnarray}

Now we are prepared to analyze the different numerical results
obtained from the direct numerical solution of the system of eqs.
(\ref{dynamics_rho}). The numerical solutions were obtained by using
a very accurate numerical method of adaptive step-size algorithm
called Bulirsch--Stoer method, which will be explained in the next
section.

We will use the numerical values $\omega_m = 0$ and $\omega_e =
-1.1$, which are of physical interest as we will discuss it
in the next section. For these particular values, the
above equations have a fixed point given by $\bar{\rho}_e =
10\bar{\rho}_m$ and
\begin{equation}
    \bar{\rho_m} = \left( \lambda \frac{10^\beta}{\sqrt{33}} \right)^{(3/2-\alpha-\beta)^{-1}} \label{NFP1}
\end{equation}

These fixed points are displayed in figures \ref{spiral_beta} and \ref{spiral_alpha}, and their loci agree remarkably well with the corresponding ones of the numerical results.

In particular, for $\alpha=0.9$, $\beta = 1.0$ and
$\lambda=1$, eq. (\ref{NFP1}) leads to the relation $\bar{\rho}_m
\approx 0.25$, which is in agreement with the corresponding locus
shown in figure \ref{spiral_alpha}. For the next attractor,
$\alpha=0.8$, $\beta = 1.0$ and $\lambda=1$, eq
(\ref{NFP1}) gives $\bar{\rho}_m \approx 0.16$, which again
perfectly agree with the numerical result shown in figure
\ref{spiral_alpha}.

Now we comeback to the linearized system of eqs. (\ref{LSE1}) and
(\ref{LSE2}). We rewrite it explicitly as the homogeneous system of
differential equations
\begin{eqnarray}
    \dot{\rho}_m &=& \left(-\sqrt(3)\left(1+\omega_m\right) \left[\frac{\bar{\rho}_m}{2\left(\bar{\rho}_m+\bar{\rho}_e\right)^{1/2}}+\left(\bar{\rho}_m+\bar{\rho}_e\right)^{1/2}\right] + \lambda\alpha\bar{\rho}_e^{\,\,\,\beta}\right)\rho_m + \nonumber \\
    & &\left(-\sqrt(3)\left(1+\omega_m\right) \left[\frac{\bar{\rho}_m}{2\left(\bar{\rho}_m+\bar{\rho}_e\right)^{1/2}}\right] + \lambda\beta\bar{\rho}_m^{\,\,\,\alpha}\right)\rho_e \\
        \dot{\rho}_e &=& \left(-\sqrt(3)\left(1+\omega_e\right) \left[\frac{\bar{\rho}_e}{2\left(\bar{\rho}_m+\bar{\rho}_e\right)^{1/2}}+\left(\bar{\rho}_m+\bar{\rho}_e\right)^{1/2}\right] - \lambda\beta\bar{\rho}_m^{\,\,\,\alpha}\right)\rho_e + \nonumber \\
    & &\left(-\sqrt(3)\left(1+\omega_e\right) \left[\frac{\bar{\rho}_e}{2\left(\bar{\rho}_m+\bar{\rho}_e\right)^{1/2}}\right] -
    \lambda\alpha\bar{\rho}_e^{\,\,\,\beta}\right)\rho_m\text{.}
\end{eqnarray}

The stability of the fixed points of the above equations depend on
the eigenvalues given by characteristic equation associated to the
system:
\begin{equation}
    k^2 - \left( a_{11} + a_{22} \right)k + \left( a_{11}a_{22} - a_{12}a_{12}
    \right)=0\text{.}
    \label{CHA}
\end{equation}

As it is well known, depending on the roots of eq. (\ref{CHA}), the
trajectories around the fixed-point $(\tilde{\rho}_m=0,\,
\tilde{\rho}_e=0)$ will be stable or unstable, (see for instance
\cite{Elsgoltz}).

For example, if all roots of the characteristic eq. (\ref{CHA})
have negative real parts, then the trivial solution $(\rho_m\,\,\,
\rho_e)^T=(0\,\,\, 0)^T$ of the linearized system and also of the
non-linear system (\ref{dynamics_rho}) is asymptotically stable. On
the other side, if at least one of the roots of eq. (\ref{CHA}) has
a positive real part then both systems have an unstable fixed point
at $(0\,\,\, 0)^T$.

For the interacting model described by eq. (\ref{dynamics_rho}),
there are five physical parameters, $w_m$, $w_e$, $\lambda$,
$\alpha$ and $\beta$, which should be chosen according to both,
physical stability properties on one side, and compatibility with
observational data on the other side. The solutions of eq.
(\ref{CHA}) were numerically evaluated for different regions of
the parameter space, and found the interesting physical region
defined by the inequalities: $0 \leq \omega_m \leq 1/3$,
$\,\,-2-\omega_m < \omega_e < -1$, $\,\,\lambda
> 1$, $\,\,0.0155 < \alpha < 0.222$ and $0.59 < \alpha < 1.02$ for
$\omega_m=0$ and $\beta > 0.8$. From it, we will considered the subregion $0 \leq \omega_m \leq 1/3$,
$\omega_e=-1.1$, $\lambda = 1$, $\alpha = 0.9$ and $\beta = 1$. In
particular, in these regions the condition of equation (\ref{elsgoltz_ineq}) holds, which
guaranties that the linear analysis of stability also apply to the
non-linear differential equation considered.

%GP and DZ
%%%%%%%%%%%%%%%%%%%%%%%%%%%%%%%%%%%%%%%%%%%%%%%%%%%%%%%%%%%%%%%%%%%%%%%%%%%%%%%%%%%%%%%
\section{Numerical analysis}\label{numerical}
%%%%%%%%%%%%%%%%%%%%%%%%%%%%%%%%%%%%%%%%%%%%%%%%%%%%%%%%%%%%%%%%%%%%%%%%%%%%%%%%%%%%%%%

In this section we present numerical results obtained by using the
Bulirsch--Stoer method to solve the non-linear coupled system of
eqs. (\ref{dynamics_rho}). The Bulirsch--Stoer method uses an
adaptive step-size control parameter, which ensures extremely high
accuracy with comparatively little extra computational effort. In
the past, this method has proven to be very accurate for solving
non-linear differential equations \cite{GP_VHC}. In addition, we
have computed the fixed point trajectories for different values of
the five physical parameters $w_m$, $w_e$, $\lambda$, $\alpha$ and
$\beta$. The corresponding trajectories within the stability region
discussed in the previous section will be shown in the figures
below.

The evolution of matter and energy densities $\rho_m$ and $\rho_e$
depends critically on the value of the parameter $w_e$. In
particular, for $w_e > -1$, the system exhibits a smooth evolution
of the densities towards the fixed point. On the other side, if $w_e
< -1$ the system shows periodic orbits around the fixed point. In
this case, a slight variation of the exponents $\alpha$ and $\beta$
leads to spiral orbits as it is shown in figures \ref{spiral_beta}
and \ref{spiral_alpha}.

\subsection{Symmetry for $w_e$ and $w_{m}$}\label{symmetry_em}

We will now consider the particular values $\alpha=\beta=1$, which
has been claimed to give the best fit to observations \cite{Ma} for
an interaction term of the form $\lambda\rho_m^\alpha\rho_e^\beta$.
As explained above, we have used the Bulirsch--Stoer method to solve
numerically \ref{dynamics_rho}.

\begin{figure}[th]
    \centering
    \includegraphics[scale=0.4]{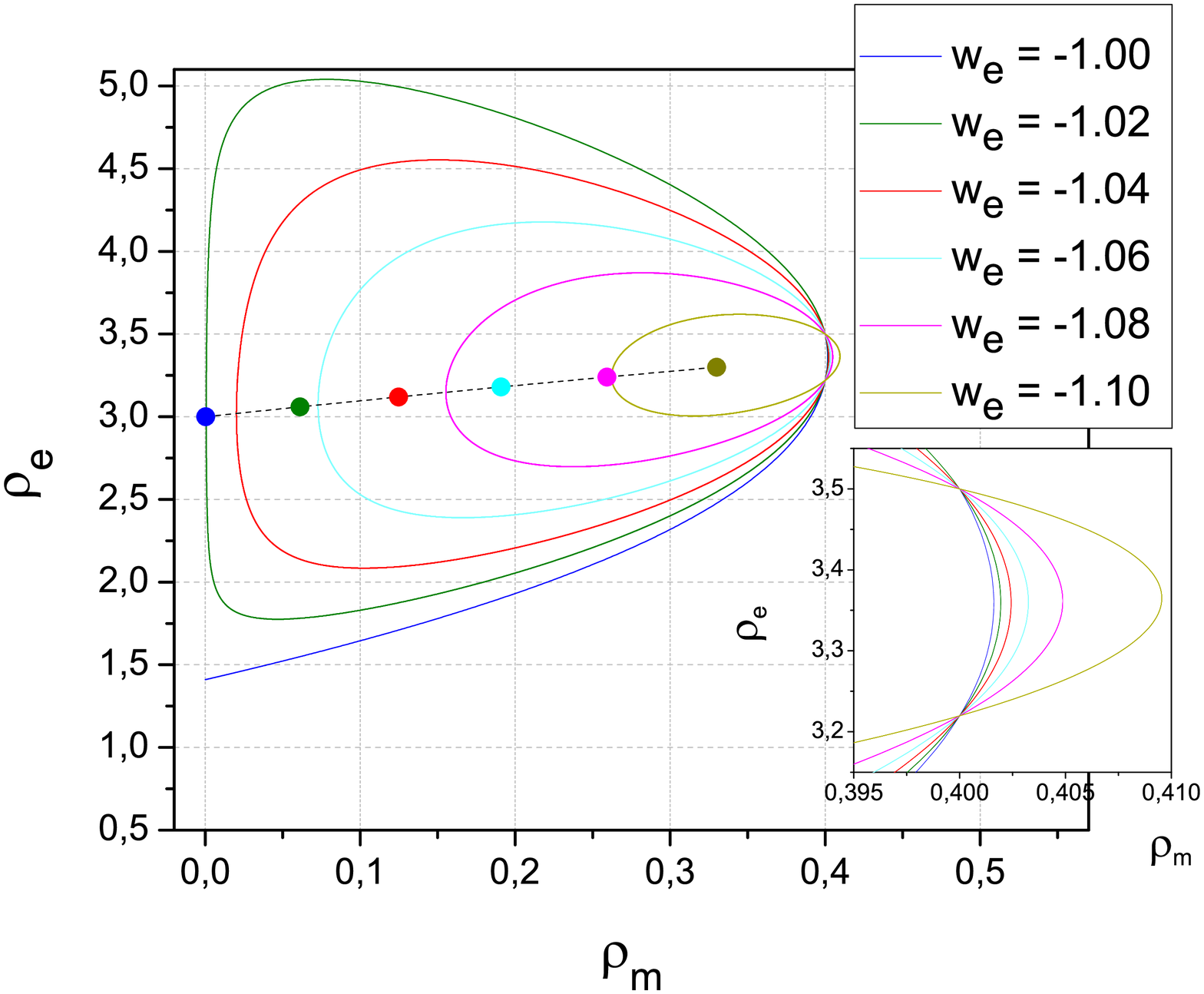}
    \caption{(Color on-line) For the parameters $\alpha=\beta=\lambda=1$,
$w_m=0$ and $w_e=[-1.0,~-1,1]$ different evolutions of the densities
are shown starting from the same initial values. Two intersection
points can be identified for the same value of $\rho_m$ when the
range of $w_e$ is swept. The dots represent the fixed points given
by eq. (\ref{Fix_P_m}) and (\ref{Fix_P_e}).}
    \label{symmetry_we.1}
\end{figure}

\begin{figure}[th]
    \centering
    \includegraphics[scale=0.4]{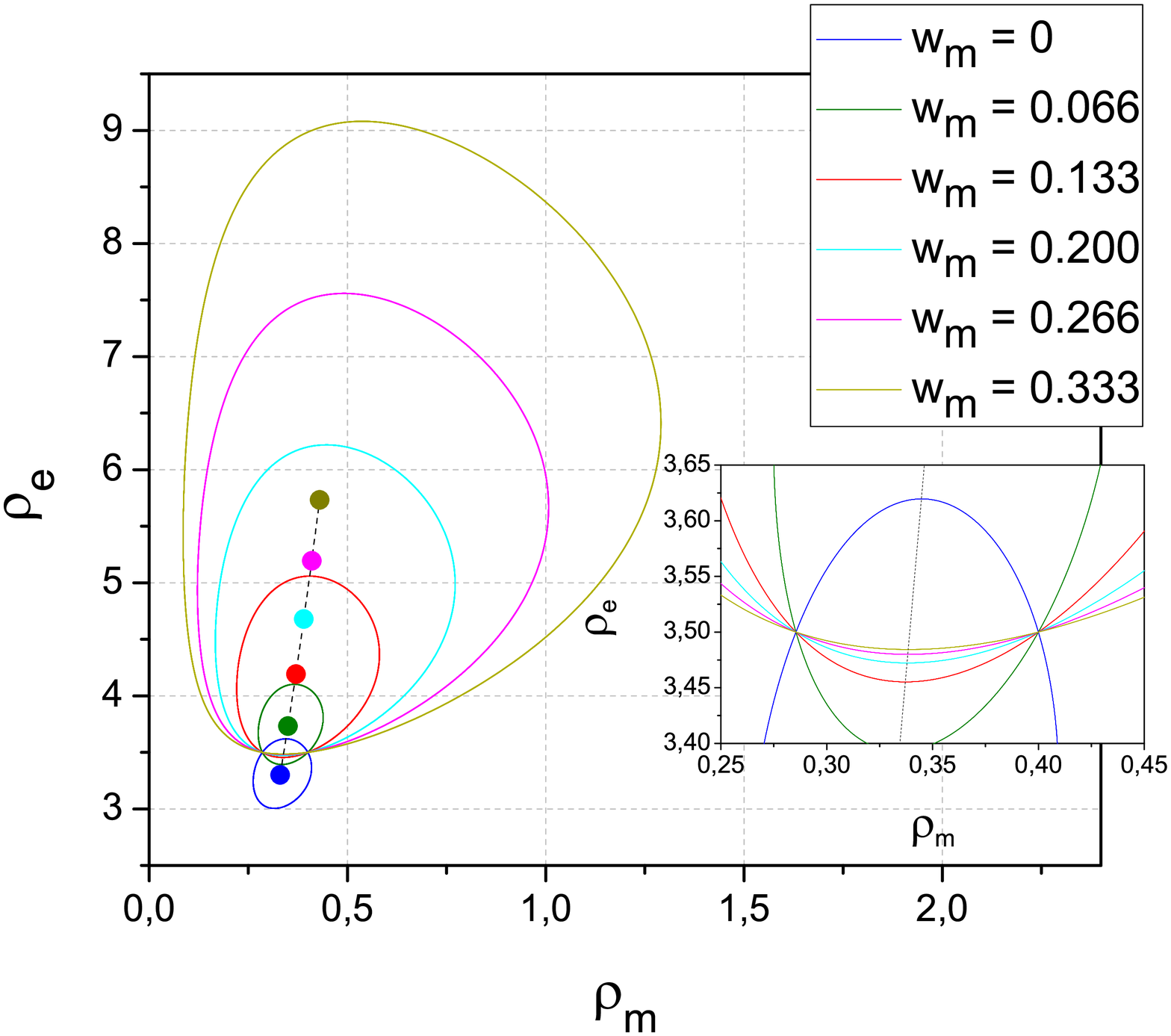}
    \caption{(Color on-line) For the parameters $\alpha=\beta=\lambda=1$,
$w_e=-1.1$ and $w_m=[0,~-1/3]$ different evolutions of the densities
are shown starting from the same initial values. Two intersection
points can be identified for the same value of $\rho_e$ when the
range of $w_m$ is swept. The dots represent the fixed points given
by eq. (\ref{Fix_P_m}) and (\ref{Fix_P_e}).}
    \label{symmetry_wm.1}
\end{figure}

For $\lambda=1$, $0<w_m<1/3$ and $-1.0<w_e<-1.1$ the trajectories of
the matter and energy densities are displayed in the figures \ref{symmetry_we.1} and \ref{symmetry_wm.1}.
For the fixed value $w_m =0$ and $w_e$ within the interval
$[-1.1,-1.0]$, different trajectories are shown in figure
\ref{symmetry_we.1}, which have the remarkable feature of having two
intersection points. These points are characterized by the same
value of $\rho_m=0.4$. For the fixed value $w_e =-1.1$ and $w_m$
within the interval $[0,1/3]$, different trajectories are shown in
figure \ref{symmetry_wm.1}. In this case the intersection points
arise at the same value of $\rho_e=3.5$.

In both cases the fixed points are of center point type, with the
exception of $w_m=0$ and $w_e=-1$, which correspond to a crossover
value (see figure \ref{symmetry_we.1}).

% CASO 3_01

% CASO 3_02

%%%%%%%%%%%%%%%%%%%%%%%%%%%%%%%%%%%%%%%%%%%%%%%%%%%%%
\subsection{Spiral trajectories}\label{spiral_tr}

\begin{figure}[th]
    \centering
    \includegraphics[scale=0.4]{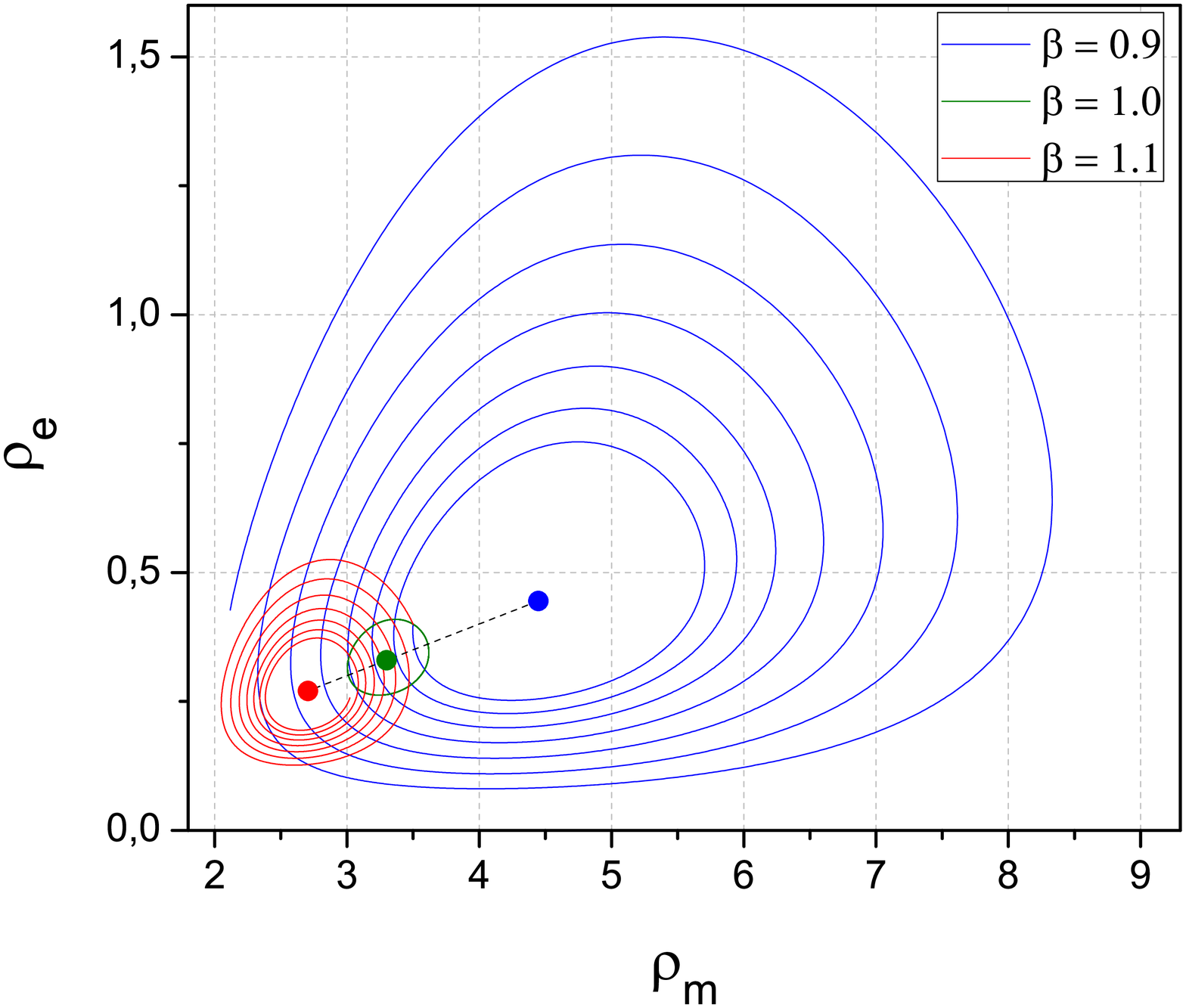}
    \caption{(Color on-line) For initial values $\rho_m = 0.4$ and $\rho_e =
3.5$ and for the parameters $\alpha=\beta=\lambda=1$, $w_m=0$ and
$w_e=-1.1$, a closed orbit is shown --blue line-- around its fixed
point --blue dot--. Starting for the same initial values but
changing the value of the power $beta$, spiral trajectories are
shown for the evolution of the densities --red and green lines--,
for $\beta=0.9$ the evolution moves away from the fixed points while
for $\beta=1.1$ the evolution is towards the fixed point.}
    \label{spiral_beta}
\end{figure}

\begin{figure}[th]
    \centering
    \includegraphics[scale=0.4]{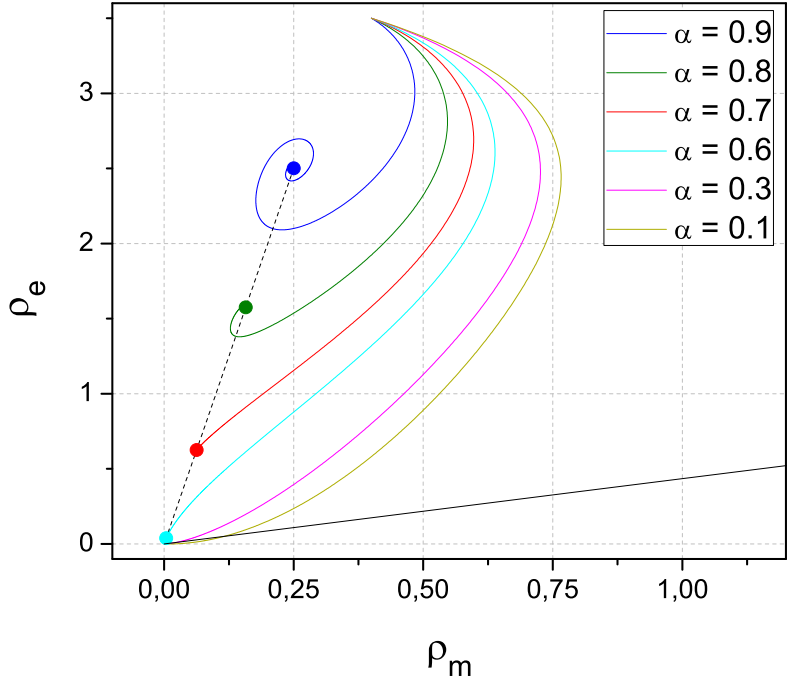}
    \caption{(Color on-line) Phase diagrams for $\rho_m$ and $\rho_e$
(starting
from the same initial conditions) are given by the color lines for different
values of $\alpha$, and for $\lambda=\beta=1$, $w_m=0$ and $w_e=-1.1$. They
were obtain by solving numerically eqs. (\ref{ICF1}) and (\ref{ICF2}). For
this figure and all the following ones, the black dashed line represent the
fixed-point trajectory given in eqs. (\ref{Fix_P_m}) and (\ref{Fix_P_e}). The
full black line is the zero-acceleration line given by the expression
(\ref{Hub2}) with $\ddot{a}=0$, the region above this line correspond to an
accelerating universe while the region below this line correspond to a
deceleration universe.}
    \label{spiral_alpha}
\end{figure}

From the previous subsection, closed trajectories are obtained only
for the region defined by $w_e<-1$. Depending on the values of
$\alpha$ and $\beta$, converging or diverging spirals are obtained.
In fact, $\alpha<1$ and $\beta>1$ lead to convergent spirals and
$\alpha>1$ and $\beta<1$ lead to divergent spirals (see figures
\ref{spiral_beta} and \ref{spiral_alpha}).

% CASO 3_03

% CASO 4_02

% data_5_0411.mat
\begin{figure}[th]
    \centering
    \includegraphics[scale=0.4]{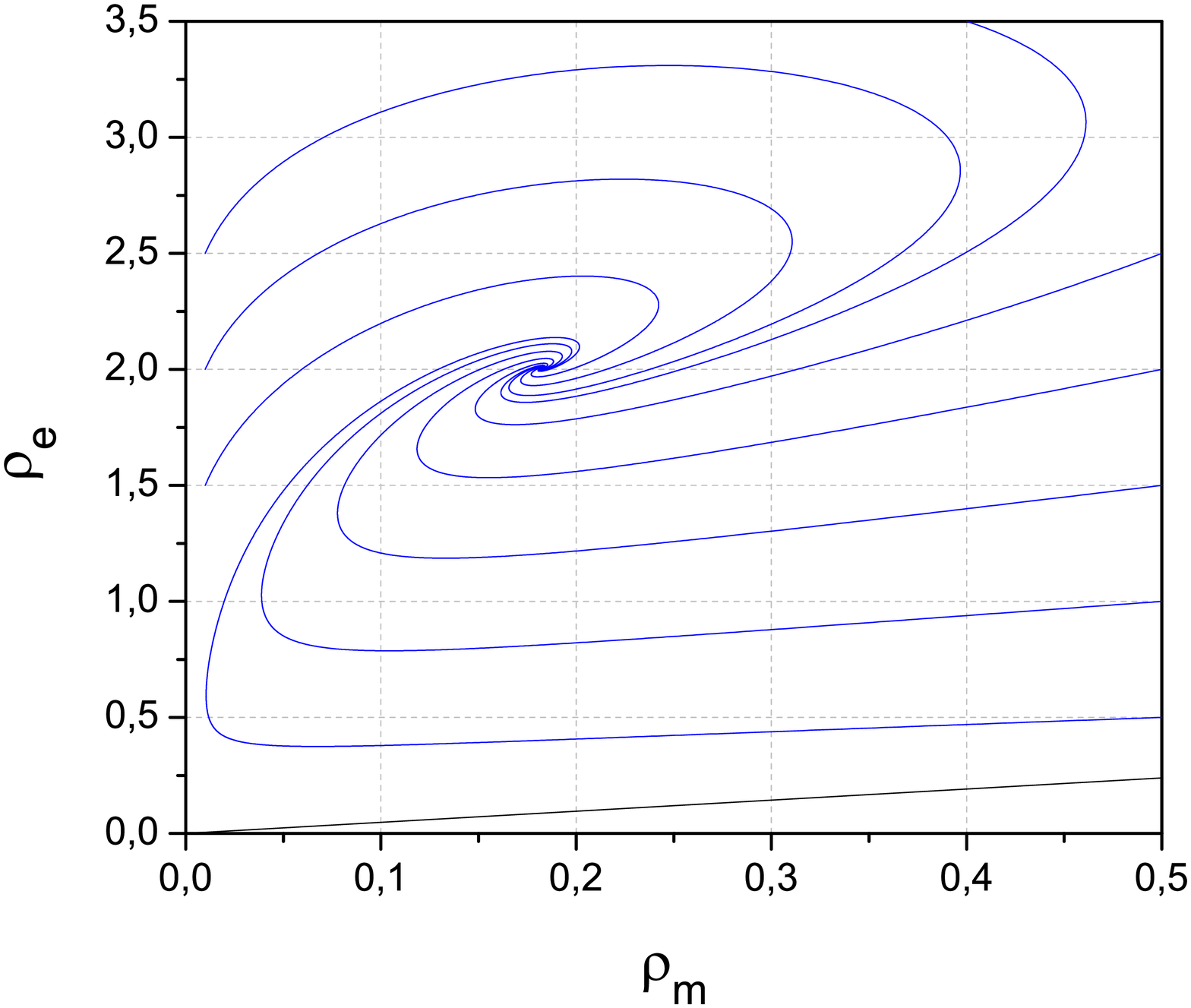}
    \caption{(Color on-line) Phase diagrams for the parameters $\alpha = 0.8$,
$\lambda=\beta=1$, $w_m=0.1$ and $w_e=-1.1$. Spiral evolution of the densities
towards the fixed points is shown.}
    \label{spiral}
\end{figure}

\begin{figure}[th]
    \centering
    \includegraphics[scale=0.5]{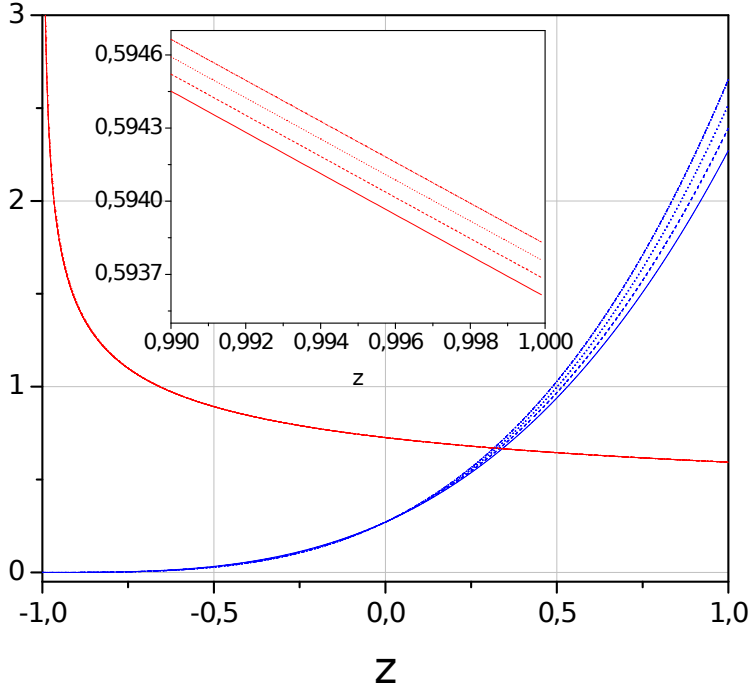}
    \caption{(Color on-line) Evolution for $\hat{\Omega}_{\rm m}(z)$ (blue)
and $\hat{\Omega}_{\rm e}(z)$ (red), with the initial conditions
$\hat{\Omega}_{\rm m}(z=0) \equiv \Omega_{\rm m0} = 0.274$ and
$\hat{\Omega}_{\rm e}(z=0) \equiv \Omega_{\rm e0} =  0.726$ (present
time $z=0$). The parameters are $\alpha = 0.9$, $\beta=1$,
$w_e=-1.1$ and $\bar{\lambda} = 1$. Full lines are for $w_m=0.025$,
dashed lines are for $w_m=0.050$, doted lines are for $w_m=0.075$
and dash-doted lines are for $w_m=0.100$. We find that the The
dependence of $\hat{\Omega}_{\rm e}(z)$ (red) with respect to $w_m$
remains almost constant compared to $\hat{\Omega}_{\rm m}(z)$ (blue)
along the interval $-1<z<1$.}
    \label{spiral2}
\end{figure}

%%%%%%%%%%% NEW  begining table %%%%%%%%%%%%%

\begin{table}
  \centering
\begin{tabular}{ c c c c c }
%\multicolumn{7}{c}{$\bar{\lambda}$}\\
\multicolumn{5}{c}{\textbf{Stability Region}} \\

\hline \hline \\[-0.3cm]
$w_{\rm m}$ & $w_{\rm e}$             & $\lambda$ & $\alpha$                         & $\beta$ \\ [0.1cm]
\hline \\[-0.3cm]

[0, 1/3]    & [-2-$w_{\rm m}$, -1]    & $>$ 1        & [0.0155, 0.222] and [0.59, 1.02] & $>$ 0.8   \\[0.2cm]

\hline
\end{tabular}
\caption{Stability region obtain from eq. (\ref{CHA}).}
\label{TableStabilityRegion}
\end{table}

%%%%%%%%%%% NEW end table %%%%%%%%%%%%%

% Arturo
%%%%%%%%%%%%%%%%%%%%%%%%%%%%%%%%%%%%%%%%%%%%%%%%%%%%%%%%%%%%%%%%%%%%%%%%%%%%%%%%%%%%%%%
\section{Observational Constraints}\label{constraints}
%%%%%%%%%%%%%%%%%%%%%%%%%%%%%%%%%%%%%%%%%%%%%%%%%%%%%%%%%%%%%%%%%%%%%%%%%%%%%%%%%%%%%%%

\subsection{The Hubble parameter}\label{Hubble_par}

In a spatially flat FRW Universe, the Hubble constraint and the
conservation equations for the matter and dark energy fluids are
given by
\begin{align}
H^2 &= \frac{8 \pi G}{3} (\rho_{\rm m} + \rho_{\rm e}),
\label{1FriedmannEq} \\
\dot{\rho}_{\rm m} + 3H(\rho_{\rm m} + p_{\rm m}) & = \lambda
\rho^{\alpha}_{\rm m} \rho^{\beta}_{\rm e}, \label{ConservationEqM1} \\
\dot{\rho}_{\rm e} + 3H(\rho_{\rm e} + p_{\rm e}) & = - \lambda
\rho^{\alpha}_{\rm m} \rho^{\beta}_{\rm e}, \label{ConservationEqE1}
\end{align}

\noindent where $\lambda$ is a constant to quantify the strength of
the interaction between the matter with the dark energy. These
equations can be written in terms of the scale factor $a$ as

\begin{subequations}\label{ConservationEq2}
\begin{align}
\frac{d \rho_{\rm m}}{da} + \frac{3}{a} \rho_{\rm m} (1+w_{\rm m}) & =
\frac{\lambda \, \rho^{\alpha}_{\rm m} \rho^{\beta}_{\rm e}}{a H },
\label{ConservationEqM2} \\
\frac{d \rho_{\rm e}}{da} + \frac{3}{a} \rho_{\rm e} (1+w_{\rm e}) & = -
\frac{\lambda \, \rho^{\alpha}_{\rm m} \rho^{\beta}_{\rm e}}{a H }.
\label{ConservationEqE2}
\end{align}
\end{subequations}

\noindent where $\lambda >0$ as mentioned above. We define the
dimensionless
parameter $\bar{\lambda}$ related with $\lambda$ as
\begin{equation}
\lambda = \frac{\bar{\lambda} \; H_0}{(\rho^0_{\rm crit})^{\alpha -1}
(\rho^0_{\rm crit})^{\beta} },
\end{equation}

\noindent where $\rho^0_{\rm crit} \equiv 3H^2_0 / (8 \pi G)$ is the
\textit{critical density} evaluated today and $H_0$ is the Hubble constant.
We define also the dimensionless parameter density $\hat{\Omega}_i
\equiv \rho_i / \rho^0_{\rm crit}$ with $i = m, e$. Using these
definitions, the Friedmann constraint equation (\ref{1FriedmannEq}) can be
expressed as $H = H_0 \sqrt{\hat{\Omega}_{\rm m} + \hat{\Omega}_{\rm e}}$,
and the conservation eqs. (\ref{ConservationEq2}) become

\begin{subequations}\label{ConservationEq3}
\begin{align}
\frac{d \hat{\Omega}_{\rm m}}{da} + \frac{3}{a} \hat{\Omega}_{\rm m}
(1+ w_{\rm m}) & = \bar{\lambda}\; \frac{
\hat{\Omega}^{\alpha}_{\rm m} \, \hat{\Omega}^{\beta}_{\rm e} }{a
\sqrt{\hat{\Omega}_{\rm m} + \hat{\Omega}_{\rm e}}},
\label{ConservationEqM3} \\
\frac{d \hat{\Omega}_{\rm e}}{da}  + \frac{3}{a} \hat{\Omega}_{\rm e}
(1+ w_{\rm e})  & = - \bar{\lambda} \; \frac{
\hat{\Omega}^{\alpha}_{\rm m} \, \hat{\Omega}^{\beta}_{\rm e} }{a
\sqrt{\hat{\Omega}_{\rm m} + \hat{\Omega}_{\rm e}}}.
\label{ConservationEqE3}
\end{align}
\end{subequations}

Using the relationship between the scale factor and the redshift $z$
given by $a= 1/(1+z)$ we express eqs. (\ref{ConservationEq3}) in
terms of the redshift as

\begin{subequations}\label{ODEs}
\begin{align} \label{ConservationEqM4z}
\frac{d \hat{\Omega}_{\rm m}}{dz} & = \frac{1}{1+z} \left[ 3(1+w_{\rm m})
\hat{\Omega}_{\rm m} - \bar{\lambda} \frac{
\hat{\Omega}^{\alpha}_{\rm m} \, \hat{\Omega}^{\beta}_{\rm e} }{
\sqrt{\hat{\Omega}_{\rm m} + \hat{\Omega}_{\rm e}}}  \right], \\
\frac{d \hat{\Omega}_{\rm e}}{dz} & = \frac{1}{1+z} \left[  3(1+w_{\rm e})
\hat{\Omega}_{\rm e} + \bar{\lambda} \frac{ \hat{\Omega}^{\alpha}_{\rm e} \,
\hat{\Omega}^{\beta}_{\rm e} }{ \sqrt{\hat{\Omega}_{\rm m} + \hat{\Omega}_{\rm
e}}} \right].
\label{ConservationEqE4z}
\end{align}
\end{subequations}

We solve numerically this ordinary differential equation system (ODEs) for
the functions $\hat{\Omega}_{\rm m}(z)$ and $\hat{\Omega}_{\rm e}(z)$, with
the initial conditions $\hat{\Omega}_{\rm m}(z=0) \equiv \Omega_{\rm m0} =
0.274$, and $\hat{\Omega}_{\rm e}(z=0) \equiv \Omega_{\rm e0} =
0.726$.

The dimensionless Hubble parameter $E \equiv H/ H_0$ becomes

\begin{equation}\label{DimensionlessHubblePar}
E(z, w_{\rm m}, w_{\rm e}, \bar{\lambda}, \alpha, \beta) =
\sqrt{ \hat{\Omega}_{\rm m}(z) + \hat{\Omega}_{\rm e}(z)},
\end{equation}

\noindent where $\hat{\Omega}_{\rm m}(z)$ and $\hat{\Omega}_{\rm e}(z)$
are given by the solution of the ODEs (\ref{ODEs}).

%%%%%%%%%%% NEW  begining table %%%%%%%%%%%%%

\begin{table}
  \centering
  \begin{tabular}{  c c c c c | c c }
%\multicolumn{7}{c}{$\bar{\lambda}$}\\

\multicolumn{6}{c}{\textbf{Best estimates,} SNe + $H(z)$} \\

\hline \hline \\[-0.3cm]

$w_{\rm m}$ & $w_{\rm e}$ & $\bar{\lambda}$ & $\alpha$ & $\beta$  &
$\chi^2_{\rm min}$ & $\chi^2_{\rm d.o.f.}$ \\ [0.1cm]
\hline \\[-0.3cm]

$0.038^{+10.5}_{-0.038}$ & $-1.017^{+1.017}_{-3.94}$ & $0.40^{+775.8}_{-0.4}$
& $0.28 \pm 1124.06$ & $0.73 \pm 10823.3$ & 573.278 & 0.97 \\[0.2cm] 

\hline

\end{tabular}

\caption{The best estimated values for the parameters $(w_{\rm m}, w_{\rm e},
\bar{\lambda}, \alpha, \beta)$, computed using the joint SNe +
$H(z)$ data sets together. The sixth and seventh columns show the
minimum of the $\chi^2$ function and its corresponding $\chi^2$
function by degrees of freedom, $\chi^2_{\rm d.o.f.}$,
defined as $\chi^2_{\rm d.o.f.} \equiv \chi^2_{\rm min}/(n-p)$ where
$n$ is the number of data ($n=592$) and $p$ the number of free
parameters estimated ($p=5$). The errors correspond to 68.3\% of
confidence level. The covariance matrix is given in expression
(\ref{CovarianceMatrix}) and the figure \ref{FigureAllCINew} shows the
marginal confidence intervals for pairs of the $(w_{\rm m}, w_{\rm e},
\bar{\lambda}, \alpha, \beta)$ parameters. The nuisance Hubble
constant $H_0$ parameter was marginalized assuming a \textit{flat}
prior probability function. From the computed value of $\chi^2_{\rm
d.o.f.} = 0.97$, we find that the model has a good fit to data. We
also find a very large dispersion on the values of the power
parameters $(\bar{\lambda}, \alpha, \beta)$, therefore, we are not able to set
stronger or useful constraints on these two parameters, both
positive and negative values for $(\alpha, \beta)$ in a a large
range are almost equally likely. For $\bar{\lambda}$ we limit ourselves to
values of $\bar{\lambda}>0$. For $w_{\rm m}$ we find a
non vanishing value as best estimate, suggesting a \textit{warm} nature for
the dark matter fluid. And for $w_{\rm e}$ the best estimated value  lies
in the phantom regime, however, given the statistical error in its estimation
we cannot be conclusive about the phantom nature of the dark energy component.
We find interesting that we obtain the stronger constraint (i.e., less
dispersion) in its value, compared to the other parameters.}
\label{TableBestEstimated}
\end{table}

%%%%%%%%%%% NEW end table %%%%%%%%%%%%%

%--------- New figures --- BEGIN
\begin{figure}
\begin{center}
\hfill
\includegraphics[width=3.6cm]{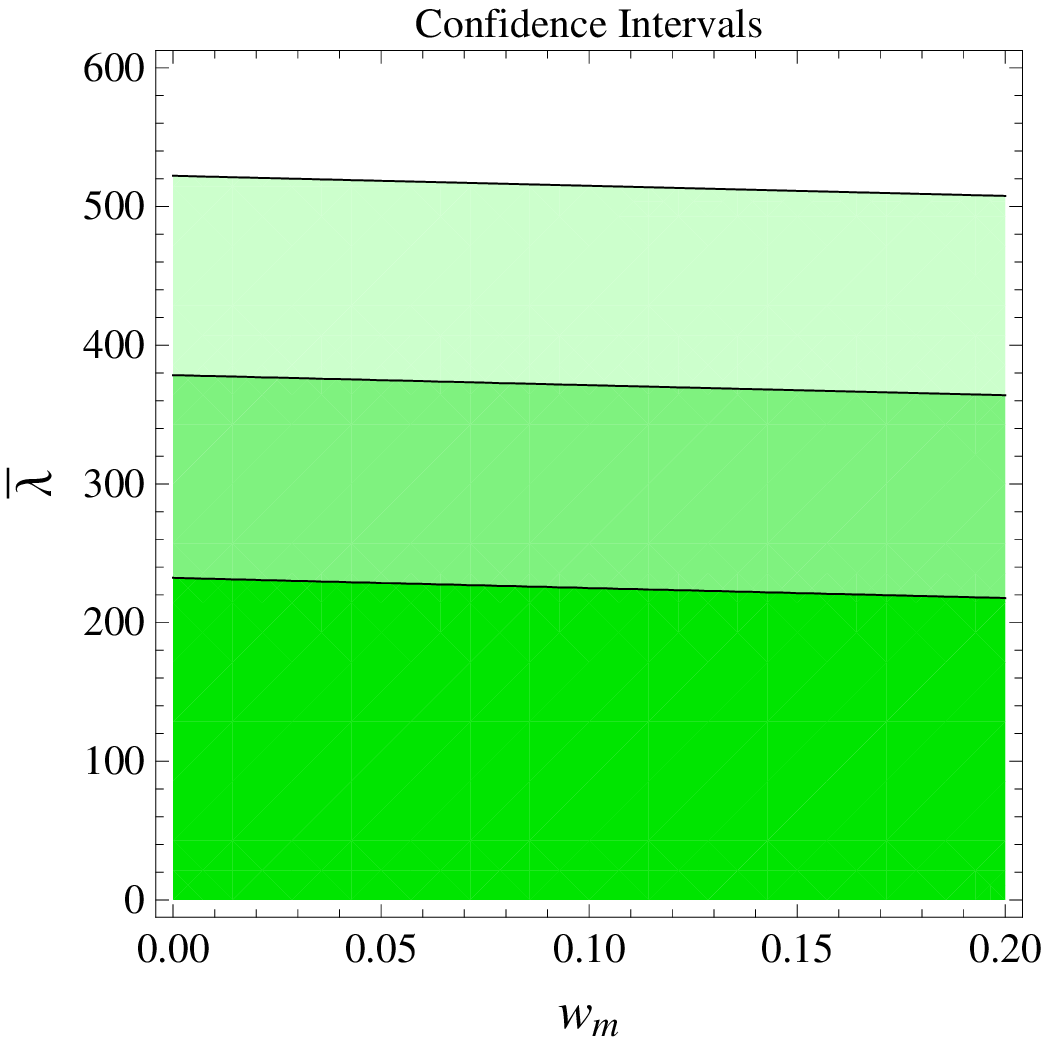}%
\hfill
\includegraphics[width=3.7cm]{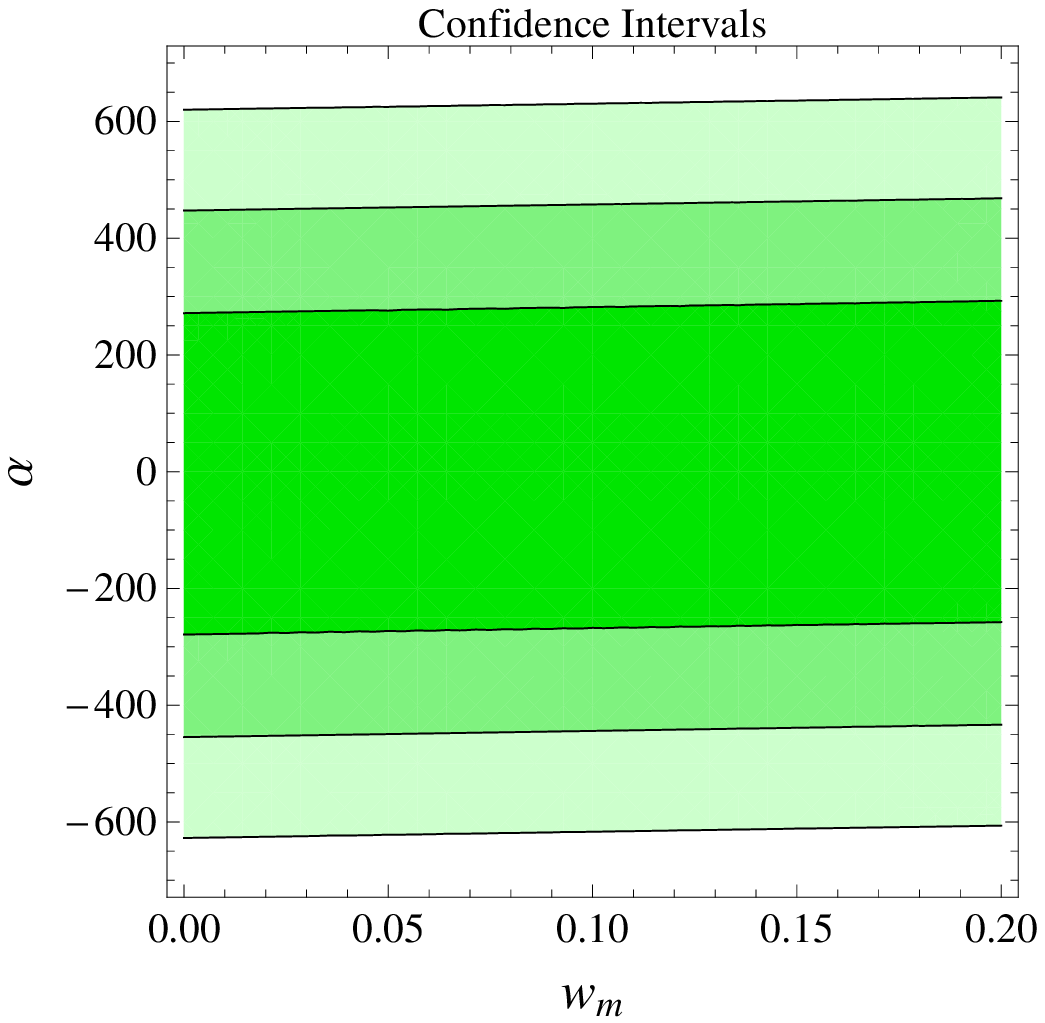}%
\hfill
\includegraphics[width=3.8cm]{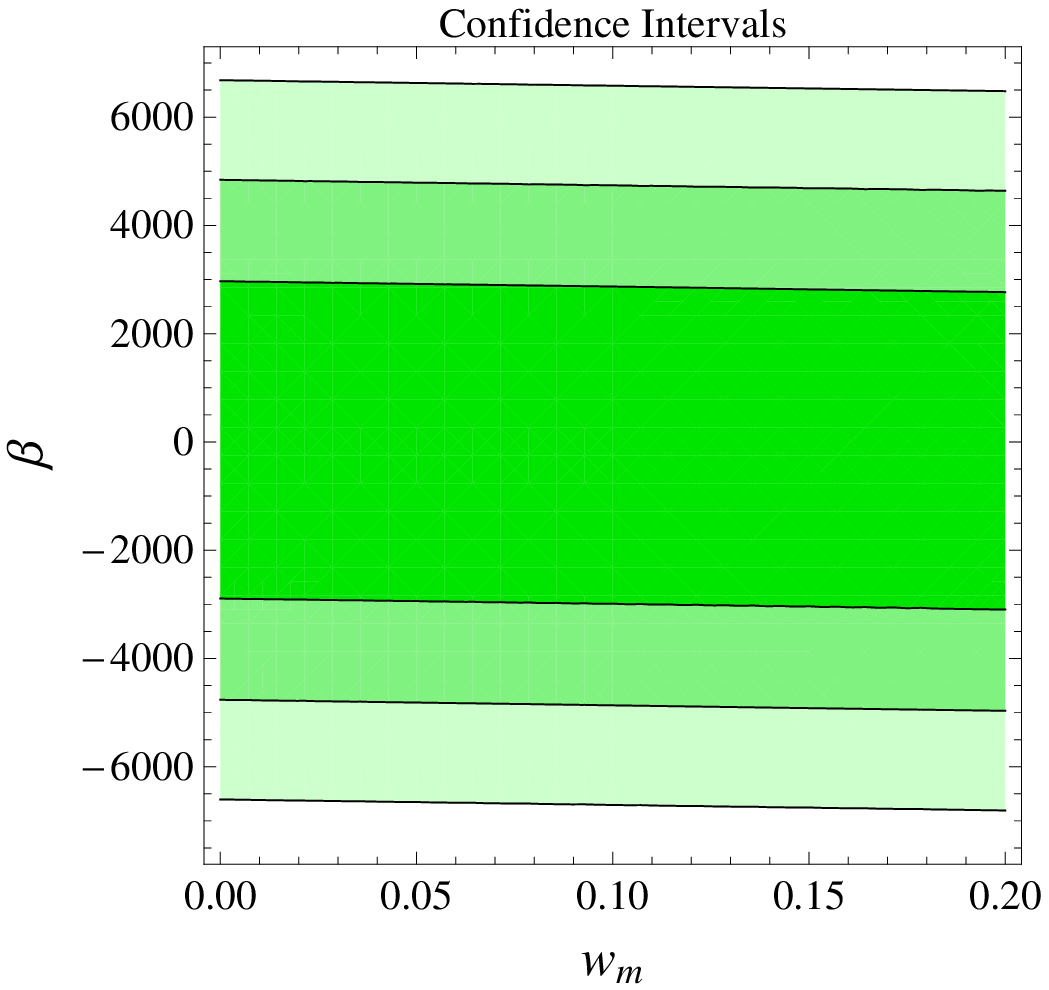}%
\hfill
\includegraphics[width=3.6cm]{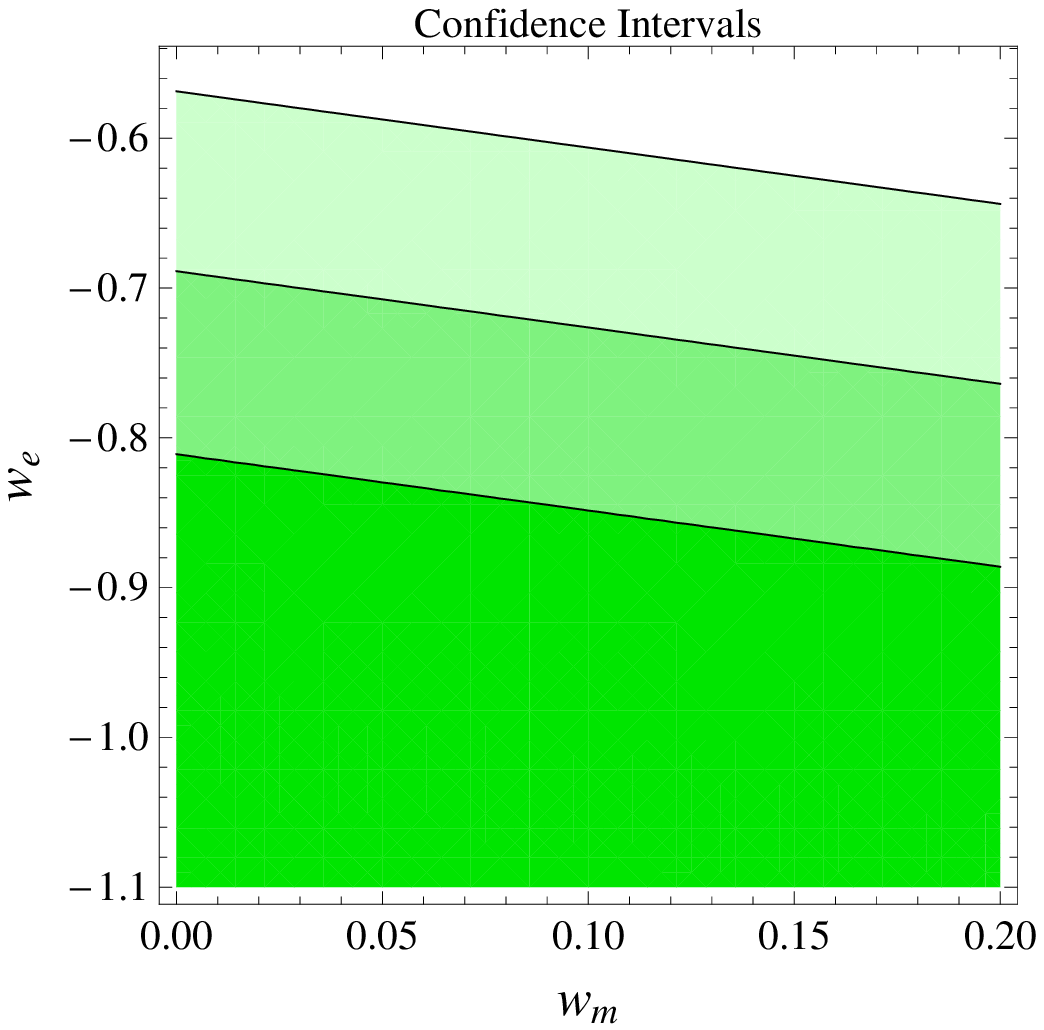}
\hspace*{\fill}
\end{center}

%\vspace{-0.4cm}
%\hspace{2.6cm} (a) \hspace{4.4cm} (b) \hspace{4.4cm} (c)
%\vspace{0.3cm}

\begin{center}
\hfill
\includegraphics[width=5cm]{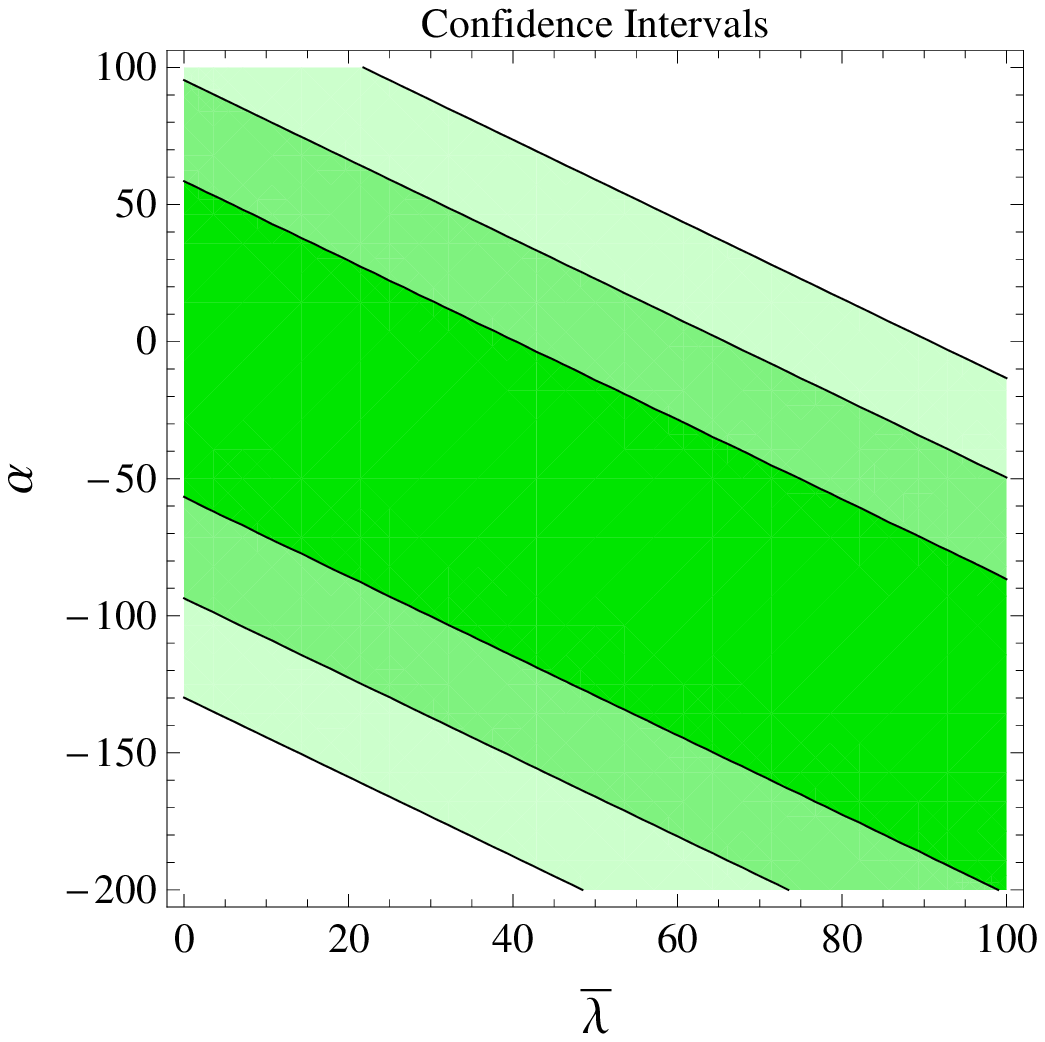}%
\hfill
\includegraphics[width=5cm]{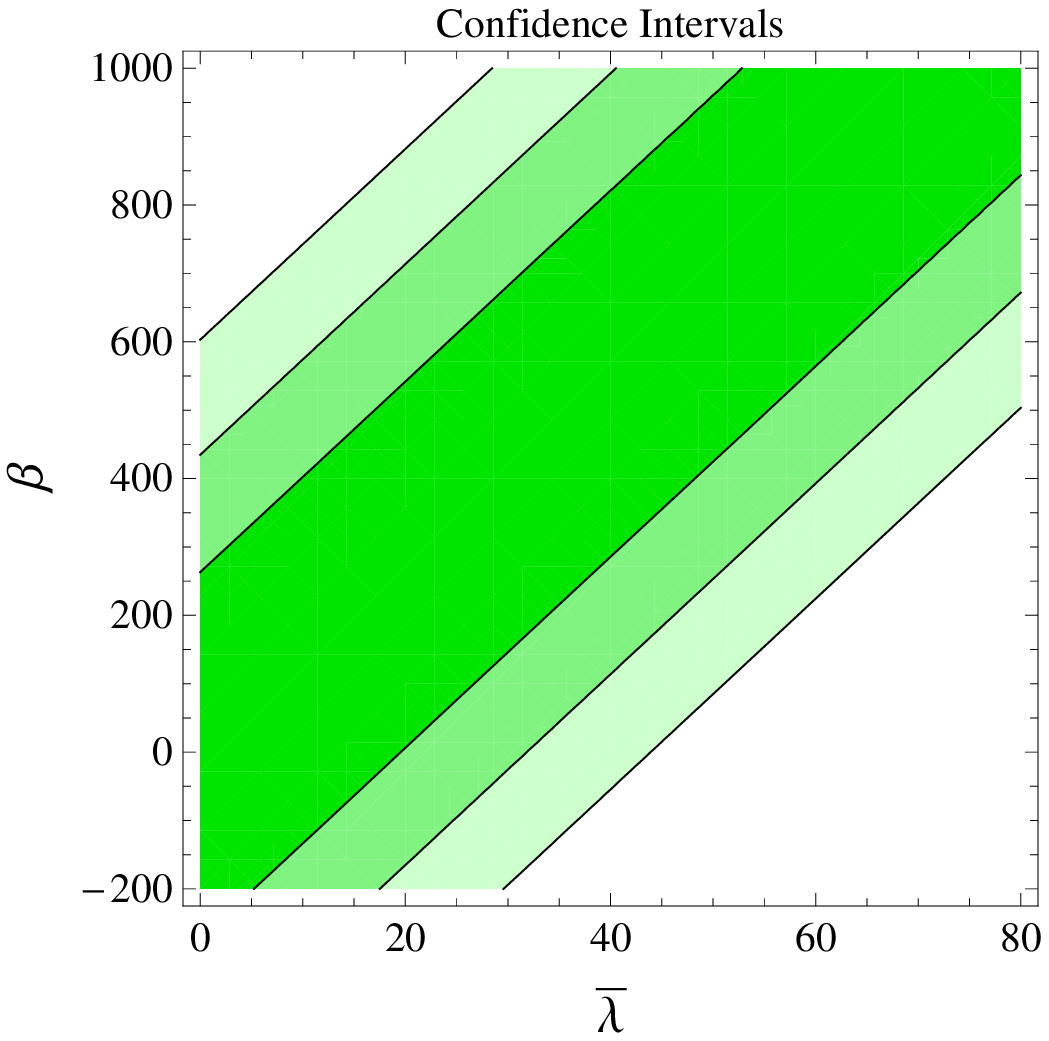}%
\hfill
\includegraphics[width=4.8cm]{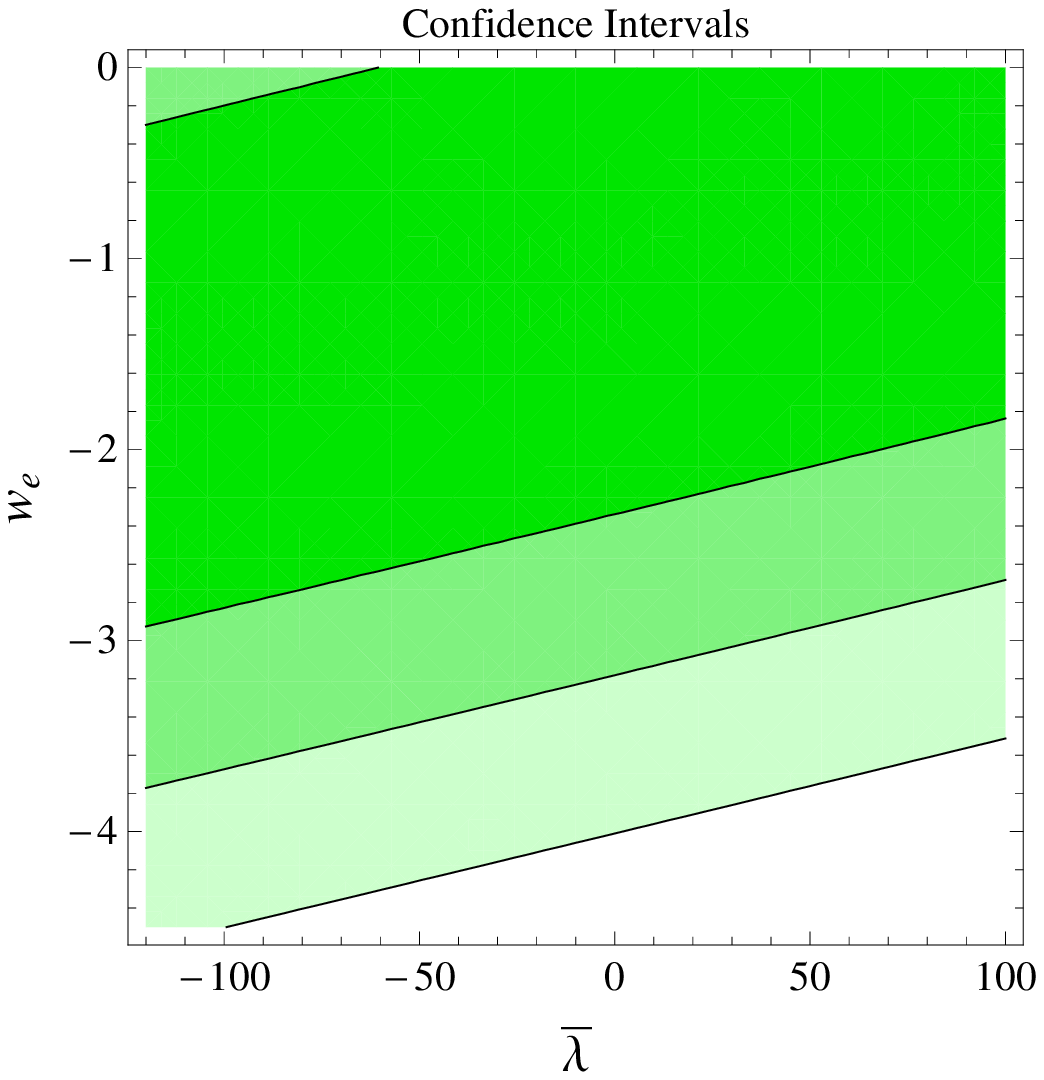}
\hspace*{\fill}
\end{center}

\begin{center}
\hfill
\includegraphics[width=5cm]{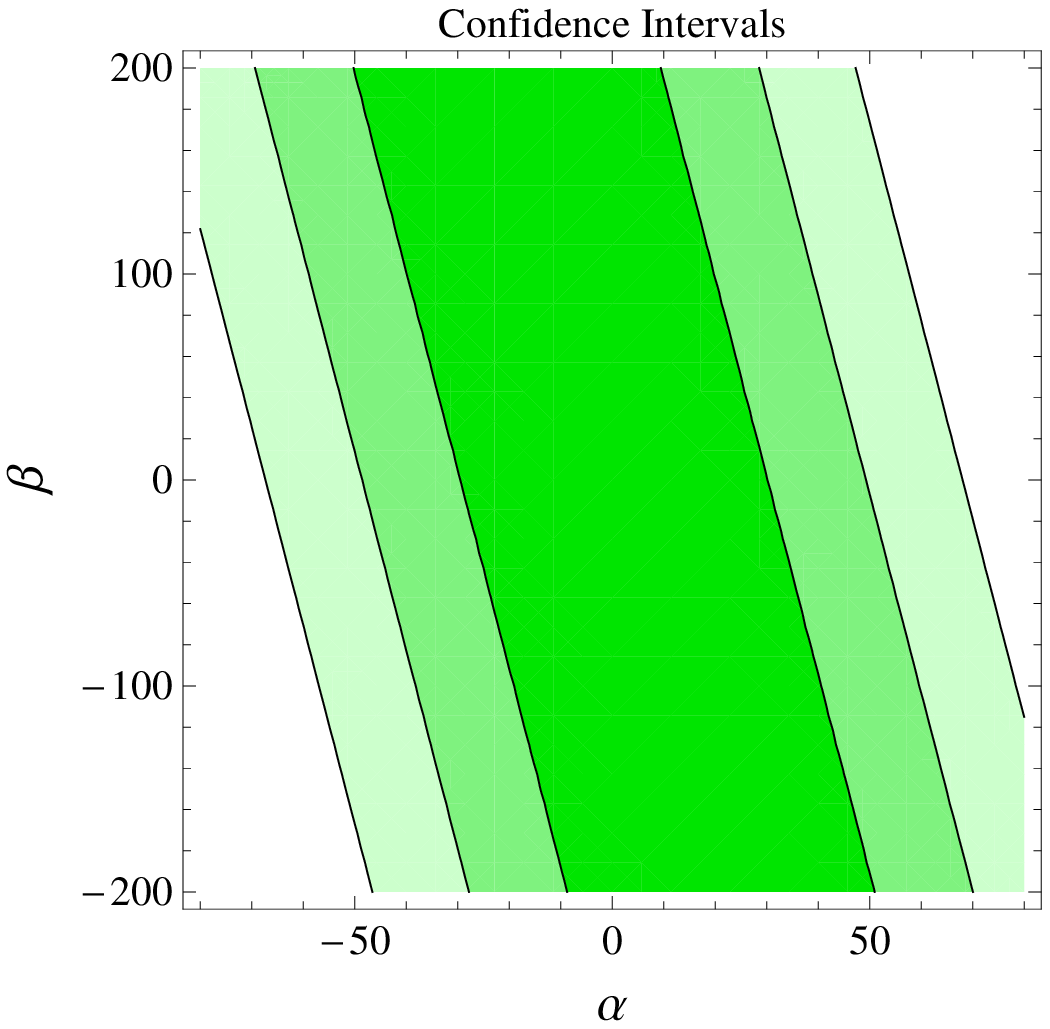}%
\hfill
\includegraphics[width=5cm]{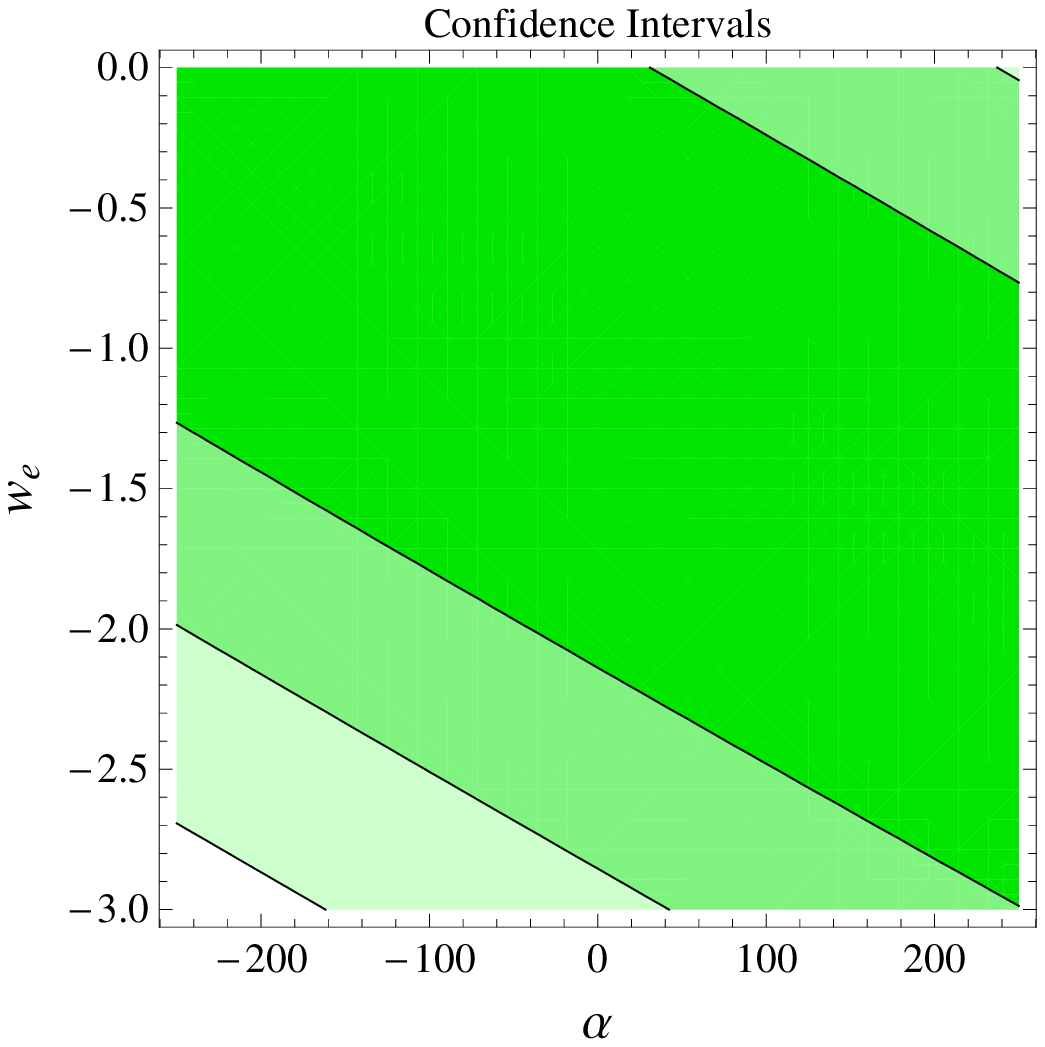}%
\hfill
\includegraphics[width=4.8cm]{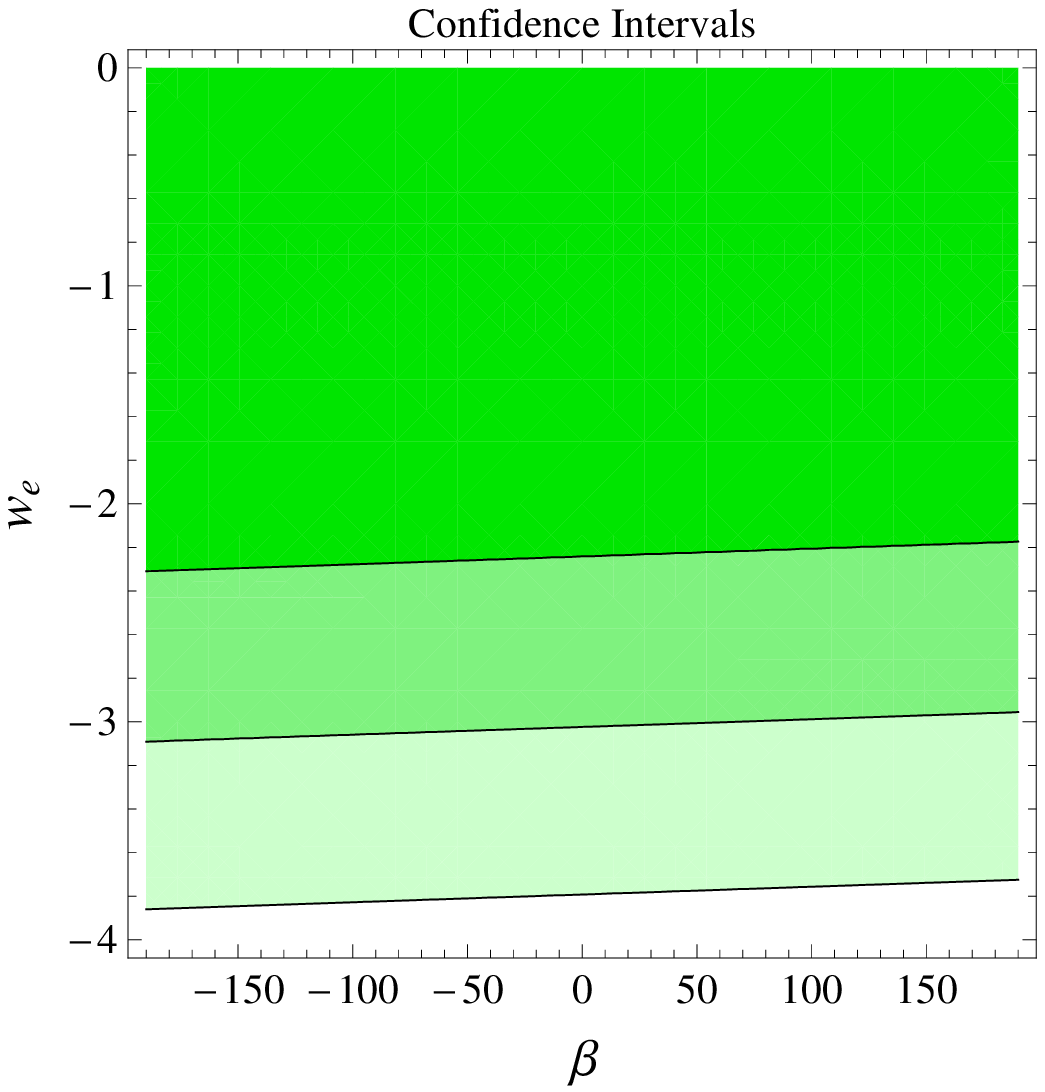}
\hspace*{\fill}
\end{center}

\caption{(Color on-line). Marginal confidence intervals (CI) from
the five parameters space $(w_{\rm m}, w_{\rm e}, \bar{\lambda}, \alpha,
\beta)$, computed together using the joint SNe + $H(z)$ data sets.
In each panel, the contour plots were computed marginalizing over
the other two remaining parameters. The CI correspond to $68.3 \%$,
$95.4 \%$ and $99.7 \%$ of confidence level. The best estimated
values for  $(w_{\rm m}, w_{\rm e}, \bar{\lambda}, \alpha, \beta)$ are shown
in table \ref{TableBestEstimated} and the covariance matrix
$\mathbf{C}$ is given in (\ref{CovarianceMatrix}). For the dark
matter barotropic index $w_m$, we find that non vanishing values are
compatible with the observations of SNe + $H(z)$.
Moreover, marginalizing over the other four parameters we find a
value of $w_{\rm m}=0.034^{+0.51}_{-0.034}$ suggesting a preference
for a \textit{warm} dark matter instead of a cold one from the
present model and data used. 
On the other hand, we notice a large
dispersion in the contour plots, in particular for the power
parameters $(\bar{\lambda},\alpha, \beta)$, indicating that a large range of
positive and negative values are allowed for both $(\alpha, \beta)$
with almost the same statistical confidence level; we are
not able to set more useful constraints on these three parameters.}
\label{FigureAllCINew}
\end{figure}

%--------- New figures --- END

\subsection{Cosmological probes}\label{cosmo_prob}
\label{SectionObservationalConstraints}

We constrain the values of  the free parameters $(w_{\rm m}, w_{\rm e},
\bar{\lambda}, \alpha, \beta)$ using cosmological
observations that measure the expansion history of the Universe, which will be explained
in the following sections. We compute their best estimated values through a
minimizing process of a $\chi^2$ function defined below, and calculate the
marginalized confidence intervals and covariance matrix of the five
parameters.

\subsubsection{Type Ia Supernovae}

We use the type Ia supernovae (SNe Ia) of the ``Union2.1'' data set
(2012) from the Supernova Cosmology Project (SCP) composed of 580 SNe
Ia \cite{SNe-Union2.1:Suzuki2011}.
The luminosity distance $d_L$ in a spatially flat FRW Universe is defined
as

\begin{equation}
d_L(z) = \frac{c(1+z)}{H_0}
\int_0^z
\frac{dz'}{E(z')}
\end{equation}

\noindent where $E(z)$ corresponds to the expression
(\ref{DimensionlessHubblePar}), and ``$c$'' to the speed of
light in units of km/sec. The theoretical distance moduli $\mu^t$ for the
k-th supernova at a distance $z_k$ is given by

\begin{equation}
\mu^t(z) = 5 \log \left[
\frac{d_L(z) }{\rm Mpc} \right] + 25
\end{equation}

So, the $\chi^2$ function for the SNe is defined as

\begin{equation}\label{Chi2FunctionSNe}
\chi^2_{\rm SNe}(w_{\rm m}, w_{\rm e}, \bar{\lambda}, \alpha, \beta) \equiv
\sum_{k=1}^n \left( \frac{\mu^t(z_k, w_{\rm m}, w_{\rm e}, \bar{\lambda},
\alpha, \beta)-
\mu_k}{\sigmạ_k} \right)^2
\end{equation}

\noindent where $\mu_k$ is the observed distance moduli of the k-th
supernova, with a standard deviation of $\sigmạ_k$ in its
measurement, and $n= 580$.

\subsubsection{Hubble expansion rate}

For the Hubble parameter $H(z)$ measured at different redshifts, we
use the 12 data listed in table 2 of Busca et al. (2012)
\cite{Busca}, where 11 data come from references
\cite{Blake1}--\cite{HubbleParameter-Riess:2011}. We assumed $H_0 = 70$ km
s$^{-1}$
Mpc$^{-1}$ for the data of Blake et al. (2011) \cite{Blake1} as Busca
et al. suggest. The $\chi^2$ function is defined as

\begin{equation}\label{Chi2Hz}
\chi^2_{\rm H}(w_{\rm m}, w_{\rm e}, \bar{\lambda} , \alpha, \beta) =
\sum_i^{12}
\left(\frac{H(z_i,w_{\rm m}, w_{\rm e}, \bar{\lambda} ,
\alpha, \beta) - H_i^{\rm obs}}{\sigma_{H i} } \right)^2
\end{equation}

\noindent where $H_i^{\rm obs}$ and $H(z_i, w_{\rm m}, w_{\rm e},
\bar{\lambda}, \alpha, \beta)= H_0 \cdot
E(z_i, w_{\rm m}, w_{\rm e}, \bar{\lambda} ,\alpha, \beta)$ are the observed
and theoretical values of $H(z)$ respectively. $E(z, w_{\rm m}, w_{\rm e},2
\bar{\lambda}, \alpha, \beta)$ is given by the expression 
(\ref{DimensionlessHubblePar}) and $\sigma_{H i}$ is the standard deviation of
each $H_i^{\rm obs}$ datum.

%----------

We construct the \textit{total} $\chi^2_{\rm t}$ function that
combine the SNe and $H(z)$ data sets together, as
\begin{equation}\label{Chi2SNeHz}
\chi^2_{\rm t} = \chi^2_{\rm SNe} + \chi^2_{\rm H},
\end{equation}

\noindent where $\chi^2_{\rm SNe}$ and $\chi^2_{\rm H}$ are given by
expressions (\ref{Chi2FunctionSNe}) and (\ref{Chi2Hz})
respectively.

We numerically minimize it to compute the \textit{best estimate values}
for the five $(w_{\rm m}, w_{\rm e}, \bar{\lambda} , \alpha, \beta)$
parameters
together, and measures the goodness-of-fit of the model to the data.
For that, we use a combination of some \textit{built-in} functions of the
\copyright Mathematica software as well as the Levenberg-Marquardt Method
described in the Numerical Recipes book \cite{NumericalRecipes}, to minimize
the $\chi^2$ function (\ref{Chi2SNeHz}).

We use also the definition of ``$\chi^2$ function by \textit{degrees
of freedom}'', $\chi^2_{\rm d.o.f.}$, defined as $\chi^2_{\rm d.o.f.}
\equiv \chi^2_{\rm min}/(n-p)$ where $n$ is the number of
\textit{total} combined data used and $p$ the number of free
parameters estimated. For our case $(n=592, p = 4)$.

The numerical results are summarized in table
\ref{TableBestEstimated} and figure \ref{FigureAllCINew}. The
computed covariance matrix that we found corresponds to

\begin{equation}\label{CovarianceMatrix}
\mathbf{C} =
\begin{pmatrix}
109.834 & -41.265 & -7975.29 & 11625.6 & -111605  \\
-41.2651 & 15.5193  & 2980.78 & -4349.28 & 41731.6  \\
-7975.29 & 2980.78 & 601931 & -871593  & 8.39609 \times 10^6  \\
11625.6  & -4349.28 & -871593 & 1.26351 \times 10^6 & -1.21642 \times 10^7  \\
-111605 & 41731.6  & 8.39609 \times 10^6 & -1.21642 \times 10^7  & 1.17145
\times 10^8 
\end{pmatrix}
\end{equation}

%%%%%%%%%%%%%%%%%%%%%%%%%%%%%%%%%%%%%%%%%%%%%%%%%%%%%%%%%%%%%%%%%%%%%%%%%%%%%%%%%%%%%%%
\section{Discussion and Conclusions}\label{conclusions}
%%%%%%%%%%%%%%%%%%%%%%%%%%%%%%%%%%%%%%%%%%%%%%%%%%%%%%%%%%%%%%%%%%%%%%%%%%%%%%%%%%%%%%%

In order to shed some light on the coincidence problem we have
explored a cosmological model composed by a dark matter fluid
interacting with a dark energy fluid. Motivated by very recent
investigations we have considered a warm dark matter. Since
non-linear interactions represent a more physical plausible
scenario for interacting fluid we studied an interaction which is
given by the term $\lambda \rho_{\rm m}^{\alpha} \rho_{\rm
e}^{\beta}$. We have found a general result which indicates 
the positive critical points of the coincidence parameter
$r=\rho_{m}/\rho_{e}$ exist if $w_{e}<-1$, independently of the
interaction chosen and the particular EoS used to describe the dark matter. We
have considered from the beginning that the energy is transferred
from dark energy to dark matter ($\lambda >0$).

%%%%%%%%%%%%%%%%

We performed an analytical analysis of the non-linear and coupled differential equations
corresponding to the continuity equations for the dark matter and dark energy
fluids. In particular we found the fixed points and their stability
properties.

Using a high precision numerical method we solved these equations and were
able not only to confirm with high accuracy the analytical results but also to
extend the solutions beyond the validity regions of the analytical analysis.

The combined method described above allowed us to compute, in the
densities space ($\rho_{m},\rho_{e}$), the behaviour of the fixed
points in terms of the parameters $\lambda, \alpha, \beta, w_{\rm m}$ and
$w_{\rm e}$. Closed orbits were found for $w_e < -1.1$ and
$\alpha=\beta=1$. If $\alpha$ or $\beta$ are different from
the unity, these closed orbits transform into spiral trajectories, evolving
towards the origin for $\alpha<1$ and $\beta=1$, and away for the fixed point for
$\alpha=1$ and $\beta>1$. This analysis allowed to constrain the parameters
in order to have physically reasonable scenarios, that is accelerated
expansion in the late time phase of the cosmic evolution and far future
evolution with finite DM and DE densities, which corresponds to spiral
trajectories propagating from the fixed points.

%%%%%%%%%%%%%%%%%%%%%%%%%%%%%%%%%%

The parameters ($w_{\rm m}, {  w_{\rm e}}, \lambda, \alpha, \beta$) were
estimated
using the cosmological observations of the Union 2.1 type Ia supernovae  and
the Hubble expansion rate $H(z)$ data sets, that measure the late time
expansion history of the Universe. A summary of these results are shown on
table \ref{TableBestEstimated} and figure \ref{FigureAllCINew}.

% w_m
For the barotropic index $w_{\rm m}$ of the EoS of the dark
matter, we found non vanishing and positive values for $w_{\rm m}$
that are well compatible with the SNe + $H(z)$ observations.
Marginalizing over the other four parameters, we found that
{  $w_{\rm m}=0.038^{+10.5}_{-0.038}$ which indicates that a \textit{warm} dark
matter is well compatible with the observations used here.
This is also in agreement with other models and
observations indicating a warm nature of the dark matter fluid.}
However, we notice a dispersion on the value of $w_{\rm m}$ larger
than the allowed by other observations.

% w_e
For the barotropic index $w_{\rm e}$ of the dark energy component, we
find that the best estimated value of $-1.017^{+1.017}_{-3.94}$ lies in the
phantom regime, however, given the magnitude of the statistical error it is
not possible to claim that the phantom nature of the dark energy component is
favoured by the observations. We can only claim that the
phantom regime is well allowed for the considered values of $w_{\rm e}$.

% LAMBDA
For the interacting coefficient $\lambda$, we defined a
dimensionless $\bar{\lambda}$ for convenience. Using the
cosmological observations it is found that the possible values for
the interacting coefficient are in the vast range of {  $0<
\bar{\lambda}< 800$ with $68.6\%$} of confidence level.  This gives us 
{  at least}
an indication of favouring the data the interaction between the dark
fluids in this model.

% ALPHA
For the power parameters $(\alpha,\beta)$, we found a large
dispersion in their values that are consistent with the SNe + $H(z)$
observations with the same confidence level, from positive to
negative values of both parameters. So, in the present work we are
not able to set strong constraints on $\alpha$ and $\beta$ but at
least we can assert that a large variety of positive and negative
values of the powers $(\alpha,\beta)$ are allowed according to the
data.

From the computed value of $\chi^2_{\rm d.o.f.}=0.97$ {  $(\chi^2_{\rm
min} = 573.278)$}, we find that the model has a well goodness-of-fit
to the SNe + $H(z)$ data.

On the other hand, we use the Bayesian Information Criterion (BIC)
\cite{BIC-Schwarz} to determine which model formed from different
cases of the values of ($w_{\rm m}, {  w_{\rm e},} \bar{\lambda}, \alpha,
\beta$)
is favoured by the observations. The value of BIC, for Gaussian
errors of the data used, is defined as
\begin{equation}
\text{BIC} = \chi^2_{\rm min} + \nu \ln N
\end{equation}
where $\nu$ and $N$ are the  number of free parameters of the model
and the number of data used respectively. The model favoured by the
observations compared to the others corresponds to that with the
smallest value of BIC, in addition to the criterion that the value
of $\chi^2_{\rm min}$ should be about or smaller to the number of
data used (in our case, $N=592$) for that model.

Computing the magnitude of the $\chi^2$ function to measure the
goodness-of-fit of data when it is evaluated at some values of
interest for ($w_{\rm m}, {  w_{\rm e}}, \bar{\lambda}, \alpha, \beta$),
as well as the corresponding value of the BIC, we find

\begin{enumerate}[label=\roman*]
\item $\chi^2_{\rm min}(w_{\rm m}=0.038, w_{\rm e}=-1.017,
\bar{\lambda}=0.40, \alpha=0.28,
\beta=0.73) = 573.278$; BIC~$=~605.196$: This case corresponds to the
present interacting model with five parameters.

\item $\chi^2_{\rm min}(0,-1, 0, 0, 0) = 581.05$; BIC $= 593.82$: This
case corresponds to the $\Lambda$CDM model; there is not interaction
between the dark components.

\item $\chi^2_{\rm min}(0.001,-1.01, 0, 0, 0) = 580.52$; BIC $= 593.28$:
This case corresponds to a warm dark matter interacting with a phantom dark
energy; there is not interaction between the dark components.

\item $\chi^2_{\rm min}(0.001,-1, 0,0,0) = 581.384$; BIC $= 594.25$: A warm
dark matter and cosmological constant model, without interaction.

%\item $\chi^2_{\rm min}(0.038,-1.01,0,0,0) = 597.02$; BIC $= 609.7$: A
%warm dark matter and phantom dark energy, where the values corresponds to the
%best estimated in the present work; without interaction.

\item $\chi^2_{\rm min}(0.038,-1.017,0.399,0,0) = 580.69$; BIC $= 599.84$:
A warm dark matter interacting with a phantom dark energy, where the values
corresponds to the best estimated in the present work. In this case the
interacting term is just the constant $\bar{\lambda}=0.399$, i.e,
$Q=$constant. 

\item $\chi^2_{\rm min}(0.038,-1, 0.39,1,1) = 577.719$; BIC $= 609.636$: A
warm dark matter interacting with a cosmological constant. The
interacting term is of the form $Q= \lambda \rho_{\rm m} \rho_{\rm
e}$. This particular case corresponds to that studied by Lip
\cite{Lip}.
\end{enumerate}

Using the BIC as a model selection criterion, we find that the model
from the above list with the smallest value of BIC, and therefore
most favored by the observations, corresponds to that composed of a
\textit{warm} dark matter and a \textit{phantom} dark energy, without
interaction. However, the difference with $\Lambda$CDM
in the BIC value is too small so that we can just conclude that the
warm dark matter -- phantom darkk energy model described above is as good as
$\Lambda$CDM model to fit the SNe$+H(z)$ data.
The interacting model (i) is not as good as the others models, despite it has
a good fit to data ($\chi_{\rm d.o.f.} = 0.97$).

Interestingly, the model that fits worst the data is the case (vi), i.e., the
model with an interacting term $Q= \lambda \rho_{\rm m} \rho_{\rm e}$. This
conclusion is in conflict with that of Lip (2011) \cite{Lip}.

In summary, from the dynamical system approach, the non linear interaction
chosen in this work leads to plausible scenarios that can alleviate the
coincidence problem. The stable fixed points represent universes which end in
a dark sector with non zero and finite energy densities in both fluids,
despite the phantom behaviour of the dark energy fluid. 

On the other hand, using the SNe + $H(z)$ observations, the best estimated
values for the free parameters of the model correspond to a warm dark matter
interacting with a phantom dark energy component, with a well goodness-of-fit
to data measured through the obtained magnitude of $\chi_{\rm d.o.f.}$.
However, using the BIC model criterion we find that this model is overcame
by a warm dark matter -- phantom dark energy model without interaction, as
well as by the $\Lambda$CDM model.

\acknowledgments
%This work was supported by CONICYT through Grant FONDECYT N$^\text{o}$ 1110840 and by DICYT-USACH Grant No 041331CM (NC).  A. A.  was partially supported through Grant FONDECYT N$^\text{o}$ 1110840 for his research visit to Departamento de F\'{\i}sica of the Universidad de Santiago de Chile.

This work was supported by CONICYT through Grant FONDECYT N$^\text{o}$ 1110840 and by DICYT-USACH grants N$^\text{o}$ 041231PA and N$^\text{o}$ 041331CM. A. A. acknowledges the very kind hospitality of Profs. G. P. and N. C. and the Departamento de F\'{\i}sica of the Universidad de Santiago de Chile where a substantial part of the work was done. We want finally to acknowledge the referee for valuable comments which contributed to an improvement of our article.

\end{document}